\documentclass[preprint,12pt]{aastex}

\usepackage{xspace}
\usepackage{natbib}
\usepackage{lscape}
\usepackage{color}

\usepackage{natbib}
\usepackage{aas_macros}
\newcommand{\cloudy}{\textsc{Cloudy}}

\font\manual=manfnt at 7pt \def\dbend{\hbox{\raise0.9ex\hbox{\manual\char127\hspace{0.6em}}}}

\newcommand\Ion[2]{\ensuremath{\mathrm{#1\,\scriptstyle #2}}}

\newcounter{INTERNALionstage}
\providecommand{\ion}[2]{
  \setcounter{INTERNALionstage}{#2}%
  \Ion{#1}{\Roman{INTERNALionstage}}}

\def\gtsim{\mathrel{\hbox{\rlap{\hbox{\lower4pt\hbox{$\sim$}}}\hbox{$>$}}}}
\def\lesssim{\mathrel{\hbox{\rlap{\hbox{\lower4pt\hbox{$\sim$}}}\hbox{$<$}}}}

\def\micron{\hbox{$\mu$m}}

\def\pcc{{\rm\thinspace cm^{-3}}}

\DeclareMathAlphabet{\vib}{OML}{cmm}{m}{it}
\newcommand*{\satellite}[1]{\textit{#1}}
\newcommand*{\xmm}{\satellite{XMM-Newton}}
\newcommand*{\chandra}{\satellite{Chandra}}

\newcommand*{\rosat}{\satellite{ROSAT}}

\newcommand{\chianti}{{\sc Chianti}}
\newcommand{\stout}{{\sc Stout}}

\newcommand{\ironXXIVline}{\ensuremath{^{57}}\ion{Fe}{24}}
\newcommand{\ironiso}{\ensuremath{^{57}\mathrm{Fe}}}

\newcommand{\ironXXIV}{\ensuremath{\mathrm{Fe}^{+23}}}

\newcommand{\ngc}{NGC~1275}
\newcommand{\virgoa}{3C~274}

\newcommand{\planck}{{\em Planck}}
\newcommand{\iso}{{\em ISO}}
\newcommand{\spitzer}{{\em Spitzer}}
\newcommand{\swift}{{\em Swift}}
\newcommand{\fermi}{{\em Fermi}}
\newcommand{\alma}{{\em ALMA}}

\newcommand{\cc}{\ensuremath{\mathrm{cm}^3}}

\newcommand{\unit}[1]{\ensuremath{\mathrm{\thinspace #1}}}

\include{figset}

\begin{document}

\title{On the Observability of Optically Thin Coronal Hyperfine Structure Lines}
\shorttitle{Coronal Hyperfine Lines}
\shortauthors{Chatzikos, Ferland, Williams, \& Fabian}

\author{M.\ Chatzikos\altaffilmark{1}, G.\ J.\ Ferland\altaffilmark{1},
	R.\ J.\ R.\ Williams\altaffilmark{2}, A.\ C.\ Fabian\altaffilmark{3} }
\email{mchatzikos@gmail.com}

\altaffiltext{1}{University of Kentucky, Lexington, KY 40506, USA}
\altaffiltext{2}{AWE plc, Aldermaston, Reading RG7 4PR, UK}
\altaffiltext{3}{Institute of Astronomy, Madingley Road, Cambridge CB3 0HA}

\begin{abstract}
We present \cloudy{} calculations for the intensity of coronal hyperfine
lines in various environments.
We model indirect collisional and radiative transitions, and quantify the
collisionally-excited line emissivity in the density-temperature phase-space.
As an observational aid, we also express the emissivity in units of that in
the 0.4--0.7\unit{keV} band.
For most hyperfine lines, knowledge of the X-ray surface brightness
and the plasma temperature is sufficient for rough estimates.
We find that the radiation fields of both Perseus~A and Virgo~A
can enhance the populations of highly ionized species within 1\unit{kpc}.
They can also enhance line emissivity within the cluster core.
This could have implications for the interpretation of spectra around bright AGN.
We find the intensity of the \ironXXIVline{} $\lambda$3.068\unit{mm}
to be about two orders of magnitude fainter than previously thought,
at $\sim$20\unit{\mu K}.
Comparably bright lines may be found in the infrared.
Finally, we find the intensity of hyperfine lines in the
Extended Orion Nebula to be low, due to the shallow sightline.
Observations of coronal hyperfine lines will likely be feasible
with the next generation of radio and sub-mm telescopes.
\end{abstract}


\section{Introduction}\label{sec:intro}

\par
Detections of hyperfine lines may provide vital clues
about the astrophysical sources they are observed in.
The most important information that may be gleaned
is the abundance of the emitting ion, and, if the line
profile is resolved, the magnitude of turbulent motions.
Reliable isotopic abundances promise to place tighter
constraints on models of primordial and stellar
nucleosynthesis, while measuring turbulent motions is
of utmost importance for the physics of galaxy clusters
\citep{Sunyaev-ICMturb-2003} and molecular clouds.
In addition, if the source kinematics are independently
known and well understood, the exact line wavelength may
provide constraints on models of atomic physics.

\par
Hyperfine transitions take place primarily through the interaction of the magnetic field
produced by the electrons and a nucleus with a non-zero magnetic moment.
All nuclei possess non-zero magnetic moments, except those with even numbers for both
protons and neutrons.
The magnetic interaction causes a splitting of energy levels with differentials that
typically correspond to radio and millimeter photons.
Note that for negative nuclear magnetic moments, the sublevel with the higher angular
momentum corresponds to the lower hyperfine level.
Due to the low energy differences, excited hyperfine levels have long lifetimes,
and vanishingly small transition probabilities.
Coupled with the low abundances of these isotopes, hyperfine lines have proven
extremely challenging to observe.


\par
To date, only a handful of hyperfine transition lines have been observed astrophysically.
The most important is the 21\unit{cm} atomic hydrogen line, which has been observed in the
Milky Way \citep[e.g.,][]{DickeyLockman1990}, as well as in other galaxies
\citep[e.g.,][]{TullyFisher1977}.
Other lines have been vigorously sought for with varying degrees of success.
The 91.6\unit{cm} atomic deuterium line has been observed near the Galactic anti-center
\citep{Rogers2007}, while the $^3$\ion{He}{2} 8.665\unit{GHz} line has been observed
in Galactic planetary nebulae \citep{Bania2007}, as well as in \ion{H}{2}
regions \citep{Bania1997}.
For reference, the observed brightness temperature of the latter is about
1\unit{mK}, and a detection with the 140-foot Green Bank telescope required
a few thousand hours of exposure \citep{Rood-He3-140ft}.

\par
Despite the observational difficulties, hyperfine transitions have been
theoretically studied in several astrophysical contexts.
\citet{SunyaevChurazov1984} studied hyperfine lines associated with highly
ionized gas, $T > 10^6$~K, in supernova remnants and cool core clusters.
They determined the primary line excitation mechanism
to be resonant scattering through the $2p$ or $2s$ levels.
They also calculated the properties of lines arising from hydrogenic, lithium-,
and sodium-like ions for all elements up to cobalt, which we use extensively
in this paper.
They predicted that the most important lines should be the hydrogenic
$^{14}$N $\lambda$5.6519\unit{mm}, and the lithium-like \ironiso{}
$\lambda$3.068\unit{mm} lines.
Recently, \citet{SunyaevDocenko2007} made refined predictions for a number of lines emanating
from supernovae remnants and other Galactic structures, and obtained best-case brightness
temperatures of order 100\unit{\mu K}.

\par
On the other hand, \citet[hereafter DSD98]{DCruzSarazin1998} examined in detail
a number of excitation processes for the \ironXXIVline{} line, and concluded that
electron impact excitation through the first excited level ($2p$) is the most
important excitation mechanism.
They parametrized the emission in the hyperfine line by considering the ratio of its intensity
to that of the X-ray doublet at 10.6\unit{\AA}, estimated to about $10^{-7}$.
A nearby cool core cluster should yield an antenna temperature of $\sim$0.5~mK if observed
with a telescope such as the NRAO 12-m Telescope \citep{Sarazin-priv-2013}.
Previously, \citet{Liang1997} had unsuccessfully attempted to detect the line in
four nearby clusters in their observations with the 22-m Morpa Telescope.

\par
In this paper, we employ the development version of the spectral synthesis
code \cloudy{} \citep{CloudyReview13} at revision r8480 to extend previous
calculations.
\cloudy{} is able to solve for the ionization structure of a plasma,
self-consistently accounting for radiative and collisional processes.

\par
Table~\ref{table:isotopes} lists the nuclear properties
of all astrophysically important isotopes, as used by the
code.
Column (1) identifies the isotope, while columns (2) and (3) list the nuclear spin,
and the nuclear magnetic moment in units of nuclear magnetons \citep{Stone2005},
respectively.
The isotopic abundance relative to hydrogen is listed in column (4).
The solar abundances of \citet{Grevesse2010}, and the isotope fractions of
\citet{Asplund2009} are employed in this work.
The latter are modified so that the deuterium-to-hydrogen ratio is primordial
\citep{Pettini2001}, and the $^{13}$C isotope fraction is 1/30.

\par
Tables~\ref{table:H-like-linelist}, \ref{table:Li-like-linelist},
and \ref{table:Na-like-linelist} present, respectively, the
hydrogenic, lithium-like, and sodium-like coronal lines of
\citet{SunyaevChurazov1984}.
Column (1) identifies the emission line, while columns (2) and (3)
give the Einstein $A$, and the transition wavelength, respectively,
as given by the reference in column (4).

\par
We improve upon previous work by accounting for impact excitations
through all levels higher than $2p$, and by considering the radiative
effects due to point source radiation fields.
These are discussed in detail in Section \ref{sec:excitation}.
In Section \ref{sec:intensity-maps} we present grids of calculations
for the hyperfine line emissivities on the density--temperature
plane.
In Section \ref{sec:predictions} we present calculations for the Perseus and
Virgo galaxy clusters, as well as the Extended Orion Nebula, and discuss
their implications in Section \ref{sec:discussion}.
We summarize in Section \ref{sec:summary}.

\begin{deluxetable}{lccc|lccc}
\tabletypesize{\rm\normalsize}
\tablecolumns{4}
\tablewidth{0pt}
\tablecaption{Isotopes of Astrophysical Interest with Finite Magnetic Moment\label{table:isotopes}}
\tablehead
{
    Isotope						&
    Spin						&
    $\mu_N$\tablenotemark{a}			&
    $\alpha_\mathrm{iso}$\tablenotemark{b}	&
    Isotope						&
    Spin						&
    $\mu_N$\tablenotemark{a}			&
    $\alpha_\mathrm{iso}$\tablenotemark{b}	\\
 (1) &
 (2) &
 (3) &
 (4) &
 (1) &
 (2) &
 (3) &
 (4)
}
\startdata

\phm{$^1$}$^{1}$H 	& 0.5 & \phm{$-$}2.792847 & 1.0000E+00		&	$^{29}$Si 	& 0.5 &       $-$0.555290 & 1.5174E$-$06\\
\phm{$^1$}$^{2}$H 	& 1.0 & \phm{$-$}0.857438 & 1.6500E$-$05	&	$^{31}$P  	& 0.5 & \phm{$-$}1.131600 & 2.5700E$-$07\\
\phm{$^1$}$^{3}$He	& 0.5 &       $-$2.127498 & 1.4127E$-$05	&	$^{33}$S  	& 1.5 & \phm{$-$}0.643821 & 1.0032E$-$07\\
\phm{$^1$}$^{6}$Li	& 1.0 & \phm{$-$}0.822047 & 8.5009E$-$13	&	$^{35}$Cl 	& 1.5 & \phm{$-$}0.821874 & 2.3947E$-$07\\
\phm{$^1$}$^{7}$Li	& 1.5 & \phm{$-$}3.256427 & 1.0350E$-$11	&	$^{37}$Cl 	& 1.5 & \phm{$-$}0.684124 & 7.6536E$-$08\\
\phm{$^1$}$^{9}$Be	& 1.5 &       $-$1.177432 & 2.4000E$-$11	&	$^{39}$K  	& 1.5 & \phm{$-$}0.391470 & 9.9653E$-$08\\
$^{10}$B  		& 3.0 & \phm{$-$}1.800645 & 9.9701E$-$11	&	$^{41}$K  	& 1.5 & \phm{$-$}0.214870 & 7.1916E$-$09\\
$^{11}$B  		& 1.5 & \phm{$-$}2.688649 & 4.0131E$-$10	&	$^{43}$Ca 	& 3.5 &       $-$1.317300 & 2.9565E$-$09\\
$^{13}$C  		& 0.5 & \phm{$-$}0.702412 & 8.9668E$-$06	&	$^{45}$Sc 	& 3.5 & \phm{$-$}4.756487 & 1.4100E$-$09\\
$^{14}$N  		& 1.0 & \phm{$-$}0.403761 & 6.7446E$-$05	&	$^{47}$Ti 	& 2.5 &       $-$0.788480 & 6.6291E$-$09\\
$^{15}$N  		& 0.5 &       $-$0.283189 & 1.5481E$-$07	&	$^{49}$Ti 	& 3.5 &       $-$1.104170 & 4.8204E$-$09\\
$^{17}$O  		& 2.5 &       $-$1.893790 & 1.8571E$-$07	&	$^{51}$V  	& 3.5 & \phm{$-$}5.148706 & 8.4889E$-$09\\
$^{19}$F  		& 0.5 & \phm{$-$}2.628868 & 3.6301E$-$08	&	$^{53}$Cr 	& 1.5 &       $-$0.474540 & 4.1520E$-$08\\
$^{21}$Ne 		& 1.5 &       $-$0.661797 & 1.8961E$-$07	&	$^{55}$Mn 	& 2.5 & \phm{$-$}3.453200 & 2.6900E$-$07\\
$^{23}$Na 		& 1.5 & \phm{$-$}2.217522 & 1.7400E$-$06	&	$^{57}$Fe 	& 0.5 & \phm{$-$}0.090764 & 6.6962E$-$07\\
$^{25}$Mg 		& 2.5 &       $-$0.855450 & 3.9801E$-$06	&	$^{59}$Co 	& 3.5 & \phm{$-$}4.627000 & 9.7702E$-$08\\
$^{27}$Al 		& 2.5 & \phm{$-$}3.641507 & 2.8200E$-$06	\\

\enddata

\tablenotetext{a}{Nuclear magnetic moment, in units of nuclear magnetons, $\mu_N$ \citep{Stone2005}.}
\tablenotetext{b}{Abundance relative to hydrogen. See text.}

\end{deluxetable}

\begin{deluxetable}{lccc|lccc}
\tabletypesize{\rm\normalsize}
\tablecolumns{8}
\tablewidth{0pt}
\tablecaption{Hydrogenic Coronal Lines\label{table:H-like-linelist}}
\tablehead
  {
    Line			&
    $A_\mathrm{ul}$		&
    $\lambda$			&
    Ref.\tablenotemark{a}	&
    Line			&
    $A_\mathrm{ul}$		&
    $\lambda$			&
    Ref.\tablenotemark{a}	\\
    		&
    (s$^{-1}$)	&
    (cm)	&
		&
    		&
    (s$^{-1}$)	&
    (cm)	&
		\\
 (1) &
 (2) &
 (3) &
 (4) &
 (1) &
 (2) &
 (3) &
 (4)
  }
\startdata

\phm{$^1$}$^{1}$\ion{H }{1}	&  2.8843E$-$15 &  \phn21.1207\phn			&  1	&	$^{29}$\ion{Si}{14}	&  1.5000E$-$06 &  \phn\phn0.0379\phn			&  2	\\
\phm{$^1$}$^{2}$\ion{H }{1}	&  4.6968E$-$17 &  \phn91.6354\phn			&  1	&	$^{31}$\ion{P }{15}	&  7.8000E$-$06 &  \phn\phn0.0151\phn			&  2	\\
\phm{$^1$}$^{3}$\ion{He}{2}	&  1.9544E$-$12 &  \phn\phn3.46194			&  1	&	$^{33}$\ion{S }{16}	&  1.2000E$-$06 &  \phn\phn0.0327\phn			&  2	\\
\phm{$^1$}$^{6}$\ion{Li}{3}	&  8.1000E$-$13 &  \phn\phn3.53\phn\phn\phn		&  2	&	$^{35}$\ion{Cl}{17}	&  4.2000E$-$06 &  \phn\phn0.0213\phn			&  2	\\
\phm{$^1$}$^{7}$\ion{Li}{3}	&  4.0000E$-$11 &  \phn\phn1\phd\phn\phn\phn\phn\phn	&  2	&	$^{37}$\ion{Cl}{17}	&  2.4000E$-$06 &  \phn\phn0.0255\phn			&  2	\\
\phm{$^1$}$^{9}$\ion{Be}{4}	&  4.2000E$-$11 &  \phn\phn1.17\phn\phn\phn		&  2	&	$^{39}$\ion{K }{19}	&  1.3000E$-$06 &  \phn\phn0.0317\phn			&  2	\\
$^{10}$\ion{B }{5}		&  5.1000E$-$10 &  \phn\phn0.447\phn\phn		&  2	&	$^{41}$\ion{K }{19}	&  2.1000E$-$07 &  \phn\phn0.0578\phn			&  2	\\
$^{11}$\ion{B }{5}		&  2.2000E$-$09 &  \phn\phn0.262\phn\phn		&  2	&	$^{43}$\ion{Ca}{20}	&  7.3000E$-$05 &  \phn\phn0.0094\phn			&  2	\\
$^{13}$\ion{C }{6}		&  4.6000E$-$10 &  \phn\phn0.3874\phn			&  3	&	$^{45}$\ion{Sc}{21}	&  4.2000E$-$03 &  \phn\phn0.00224			&  2	\\
$^{14}$\ion{N }{7}		&  2.0000E$-$10 &  \phn\phn0.56519			&  3	&	$^{47}$\ion{Ti}{22}	&  4.5000E$-$05 &  \phn\phn0.0111\phn			&  2	\\
$^{15}$\ion{N }{7}		&  3.6000E$-$10 &  \phn\phn0.603\phn\phn		&  2	&	$^{49}$\ion{Ti}{22}	&  1.0000E$-$04 &  \phn\phn0.00835			&  2	\\
$^{17}$\ion{O }{8}		&  6.2000E$-$08 &  \phn\phn0.10085			&  3	&	$^{51}$\ion{V }{23}	&  1.2000E$-$02 &  \phn\phn0.00156			&  2	\\
$^{19}$\ion{F }{9}		&  9.6000E$-$07 &  \phn\phn0.0305\phn			&  2	&	$^{53}$\ion{Cr}{24}	&  3.3000E$-$05 &  \phn\phn0.0127\phn			&  2	\\
$^{21}$\ion{Ne}{10}		&  2.8000E$-$08 &  \phn\phn0.132\phn\phn		&  2	&	$^{55}$\ion{Mn}{25}	&  8.9000E$-$05 &  \phn\phn0.00171			&  2	\\
$^{23}$\ion{Na}{11}		&  1.6000E$-$06 &  \phn\phn0.0296\phn			&  2	&	$^{57}$\ion{Fe}{26}	&  6.4000E$-$07 &  \phn\phn0.0348\phn			&  2	\\
$^{25}$\ion{Mg}{12}		&  2.2000E$-$07 &  \phn\phn0.0654\phn			&  2	&	$^{59}$\ion{Co}{27}	&  4.1000E$-$02 &  \phn\phn0.00105			&  2	\\
$^{27}$\ion{Al}{13}		&  2.5000E$-$05 &  \phn\phn0.0121\phn			&  2	\\

\enddata

\tablenotetext{a}
{
	Wavelength and transition probability reference:
	1: \citet{Gould1994};
	2: \citet{SunyaevChurazov1984}; 
	3: \citet{Shabaev1995}.
}

\end{deluxetable}

\begin{deluxetable}{lccc|lccc}
\tabletypesize{\rm\normalsize}
\tablecolumns{8}
\tablewidth{0pt}
\tablecaption{Lithium-like Coronal Lines\label{table:Li-like-linelist}}
\tablehead
  {
    Line			&
    $A_\mathrm{ul}$		&
    $\lambda$			&
    Ref.\tablenotemark{a}	&
    Line			&
    $A_\mathrm{ul}$		&
    $\lambda$			&
    Ref.\tablenotemark{a}	\\
    		&
    (s$^{-1}$)	&
    (cm)	&
		&
    		&
    (s$^{-1}$)	&
    (cm)	&
		\\
 (1) &
 (2) &
 (3) &
 (4) &
 (1) &
 (2) &
 (3) &
 (4)
  }
\startdata

\phm{$^1$}$^{6}$\ion{Li}{1}	&  1.5900E$-$17 & 131.461\phn\phn		&  1	&	$^{31}$\ion{P }{13}	&  8.6000E$-$09 & \phn\phn0.146\phn\phn			&  2		\\
\phm{$^1$}$^{7}$\ion{Li}{1}	&  7.7900E$-$16 & \phn37.3364\phn		&  1	&	$^{33}$\ion{S }{14}	&  1.3000E$-$09 & \phn\phn0.312\phn\phn			&  2		\\
\phm{$^1$}$^{9}$\ion{Be}{2}	&  4.8900E$-$15 & \phn23.999\phn\phn		&  1	&	$^{35}$\ion{Cl}{15}	&  5.0000E$-$09 & \phn\phn0.20073			&  3		\\
$^{10}$\ion{B }{3}		&  1.2000E$-$13 & \phn\phn7.2\phn\phn\phn\phn	&  2	&	$^{37}$\ion{Cl}{15}	&  2.9000E$-$09 & \phn\phn0.241\phn\phn			&  2		\\
$^{11}$\ion{B }{3}		&  5.4000E$-$13 & \phn\phn4.22\phn\phn\phn	&  2	&	$^{39}$\ion{K }{17}	&  1.6000E$-$09 & \phn\phn0.294\phn\phn			&  2		\\
$^{13}$\ion{C }{4}		&  1.7000E$-$13 & \phn\phn5.39\phn\phn\phn	&  2	&	$^{41}$\ion{K }{17}	&  2.6000E$-$10 & \phn\phn0.535\phn\phn			&  2		\\
$^{14}$\ion{N }{5}		&  9.8000E$-$14 & \phn\phn7.072\phn\phn		&  3	&	$^{43}$\ion{Ca}{18}	&  9.4000E$-$08 & \phn\phn0.0862\phn			&  2		\\
$^{15}$\ion{N }{5}		&  1.8000E$-$13 & \phn\phn7.65\phn\phn\phn	&  2	&	$^{45}$\ion{Sc}{19}	&  5.6000E$-$06 & \phn\phn0.0203\phn			&  2		\\
$^{17}$\ion{O }{6}		&  3.6000E$-$11 & \phn\phn1.19\phn\phn\phn	&  2	&	$^{47}$\ion{Ti}{20}	&  6.2000E$-$08 & \phn\phn0.101\phn\phn			&  2		\\
$^{19}$\ion{F }{7}		&  6.6000E$-$10 & \phn\phn0.34102		&  3	&	$^{49}$\ion{Ti}{20}	&  1.4000E$-$07 & \phn\phn0.0754\phn			&  2		\\
$^{21}$\ion{Ne}{8}		&  2.3000E$-$11 & \phn\phn1.43\phn\phn\phn	&  2	&	$^{51}$\ion{V }{21}	&  1.7000E$-$05 & \phn\phn0.014\phn\phn			&  2		\\
$^{23}$\ion{Na}{9}		&  1.3000E$-$09 & \phn\phn0.30924		&  3	&	$^{53}$\ion{Cr}{22}	&  4.7000E$-$08 & \phn\phn0.113\phn\phn			&  2		\\
$^{25}$\ion{Mg}{10}		&  2.1000E$-$10 & \phn\phn0.6679\phn		&  3	&	$^{55}$\ion{Mn}{23}	&  1.3000E$-$05 & \phn\phn0.0152\phn			&  2		\\
$^{27}$\ion{Al}{11}		&  2.5000E$-$08 & \phn\phn0.1206\phn		&  3	&	$^{57}$\ion{Fe}{24}	&  9.4000E$-$10 & \phn\phn0.3068\phn			&  3		\\
$^{29}$\ion{Si}{12}		&  1.5000E$-$09 & \phn\phn0.3725\phn		&  3	&	$^{59}$\ion{Co}{25}	&  6.1000E$-$05 & \phn\phn0.00915			&  2		\\

\enddata

\tablenotetext{a}
{
	Wavelength and transition probability reference:
	1: \citet{Garstang1995}.
	2: \citet{SunyaevChurazov1984}; 
	3: \citet{Shabaev1995};
}

\end{deluxetable}

\begin{deluxetable}{lccc|lccc}
\tabletypesize{\rm\normalsize}
\tablecolumns{8}
\tablewidth{0pt}
\tablecaption{Sodium-like Coronal Lines\label{table:Na-like-linelist}}
\tablehead
  {
    Line			&
    $A_\mathrm{ul}$		&
    $\lambda$			&
    Ref.\tablenotemark{a}	&
    Line			&
    $A_\mathrm{ul}$		&
    $\lambda$			&
    Ref.\tablenotemark{a}	\\
    		&
    (s$^{-1}$)	&
    (cm)	&
		&
    		&
    (s$^{-1}$)	&
    (cm)	&
		\\
 (1) &
 (2) &
 (3) &
 (4) &
 (1) &
 (2) &
 (3) &
 (4)
  }
\startdata

$^{23}$\ion{Na}{1} 		& 8.3500E$-$15 &  \phn16.9336\phn			& 1	&	$^{43}$\ion{Ca}{10}	& 2.7000E$-$10 &  \phn\phn0.61\phn\phn\phn		& 2		\\
$^{25}$\ion{Mg}{2} 		& 1.3400E$-$14 &  \phn16.771\phn\phn			& 1	&	$^{45}$\ion{Sc}{11} 	& 1.8000E$-$08 &  \phn\phn0.14\phn\phn\phn		& 2		\\
$^{27}$\ion{Al}{3} 		& 5.5000E$-$12 &  \phn\phn2\phd\phn\phn\phn\phn\phn	& 2	&	$^{47}$\ion{Ti}{12} 	& 2.3000E$-$10 &  \phn\phn0.65\phn\phn\phn		& 2		\\
$^{29}$\ion{Si}{4} 		& 6.3000E$-$13 &  \phn\phn5\phd\phn\phn\phn\phn\phn	& 2	&	$^{49}$\ion{Ti}{12} 	& 5.2000E$-$10 &  \phn\phn0.49\phn\phn\phn		& 2		\\
$^{31}$\ion{P }{5} 		& 6.5000E$-$12 &  \phn\phn1.6\phn\phn\phn\phn		& 2	&	$^{51}$\ion{V }{13} 	& 7.0000E$-$08 &  \phn\phn0.087\phn\phn			& 2		\\
$^{33}$\ion{S }{6} 		& 1.3000E$-$12 &  \phn\phn3.1\phn\phn\phn\phn		& 2	&	$^{53}$\ion{Cr}{14} 	& 2.1000E$-$10 &  \phn\phn0.68\phn\phn\phn		& 2		\\
$^{35}$\ion{Cl}{7} 		& 7.0000E$-$12 &  \phn\phn1.8\phn\phn\phn\phn		& 2	&	$^{55}$\ion{Mn}{15} 	& 6.6000E$-$08 &  \phn\phn0.089\phn\phn			& 2		\\
$^{37}$\ion{Cl}{7} 		& 4.4000E$-$12 &  \phn\phn2.1\phn\phn\phn\phn		& 2	&	$^{57}$\ion{Fe}{16} 	& 5.0000E$-$12 &  \phn\phn1.8\phn\phn\phn\phn		& 2		\\
$^{39}$\ion{K }{9} 		& 3.8000E$-$12 &  \phn\phn2.2\phn\phn\phn\phn		& 2	&	$^{59}$\ion{Co}{17} 	& 3.5000E$-$07 &  \phn\phn0.051\phn\phn			& 2		\\
$^{41}$\ion{K }{9} 		& 6.2000E$-$13 &  \phn\phn4\phd\phn\phn\phn\phn\phn	& 2	&	\\

\enddata

\tablenotetext{a}
{
	Wavelength and transition probability reference:
	1: \citet{Garstang1995};
	2: \citet{SunyaevChurazov1984}.
}

\end{deluxetable}

\section{Line Excitation Processes}\label{sec:excitation}

\subsection{Collisional Excitations}\label{sec:collisional-excitation}

\subsubsection{Direct and Resonant Excitations}\label{sec:resonant-excitation}

\par
\citet{SunyaevChurazov1984} examined a number of line excitation processes
and concluded that resonant excitation through the first level above ground
is the primary excitation mechanism for the coronal hyperfine lines of
Tables~\ref{table:H-like-linelist}--\ref{table:Na-like-linelist}.
%
At the suggestion of DSD98, \citet[][hereafter ZS00]{ZhangSampson2000, ZhangSampson2001}
performed detailed calculations of the collision strength for direct and resonant excitation
of hyperfine lines for a number of hydrogenic and lithium-like ions.
Their findings suggest that resonant excitations indeed dominate direct excitations at
temperatures below $\sim 5\times10^7\unit{K}$.
In general, resonant excitations are more important than direct excitations.

\par
\citet{Goddard2003} provided fits to these collisional data of about 30\%
accuracy for the purposes of incorporating them into \cloudy{}.
In Appendix~\ref{sec:app:fitsZS} we present improved fits to the collision
strengths of Zhang \& Sampson of accuracy about 15\%, which we use
throughout our calculations.
Note that we adopt the lithium-like fits for the sodium-like collision strengths,
due to lack of data for sodium-like ions.

\par
In the following, we will refer to direct and resonant
excitations collectively as ``direct'' excitations.

\subsubsection{Indirect Impact Excitations}\label{sec:indirect-impact-excitations}

\par
On the other hand, DSD98 showed that the most important excitation mechanism
for the lithium-like \ironiso{} line is indirect impact collisional excitations
through the $2p$ sublevel.
DSD98 also considered excitations through higher levels and concluded that the
effect was not significant, of order 10\%.

\par
We have adapted \cloudy{} to include excitations through levels higher
than $2p$.
These are drawn from the \chianti{}
\citep{Dere.K97CHIANTI---an-atomic-database-for-emission, Chianti7.1},
\stout{} (Ferland et al., in preparation), and Opacity Project
\citep{OpacityProject} databases.
Note that no radiative or collisional data are available for Cl$^{+14}$.
Our results on the lithium-like \ion{Cl}{15} lines include only direct
excitations.

\par
\begin{landscape}
\begin{figure*}
	\begin{centering}
		\includegraphics[scale=0.625]{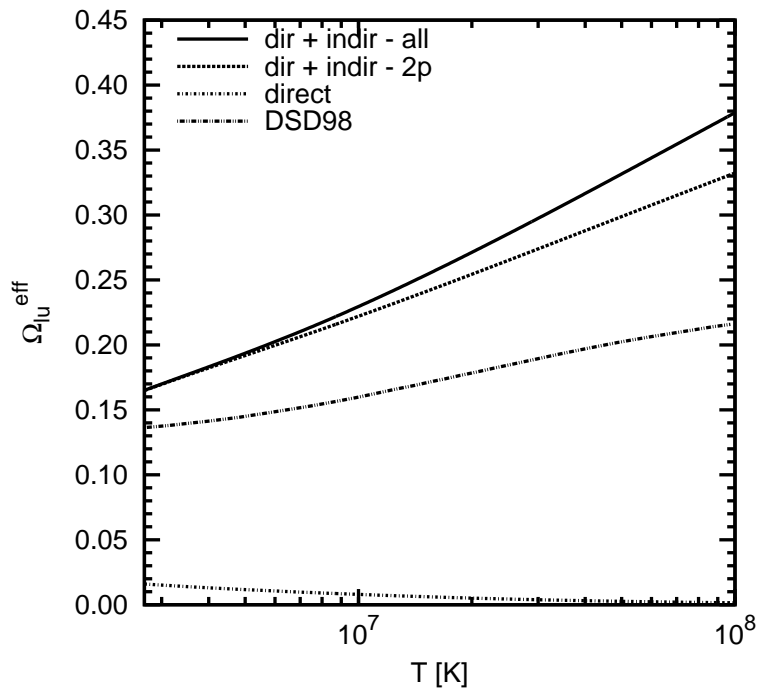}
		\includegraphics[scale=0.625]{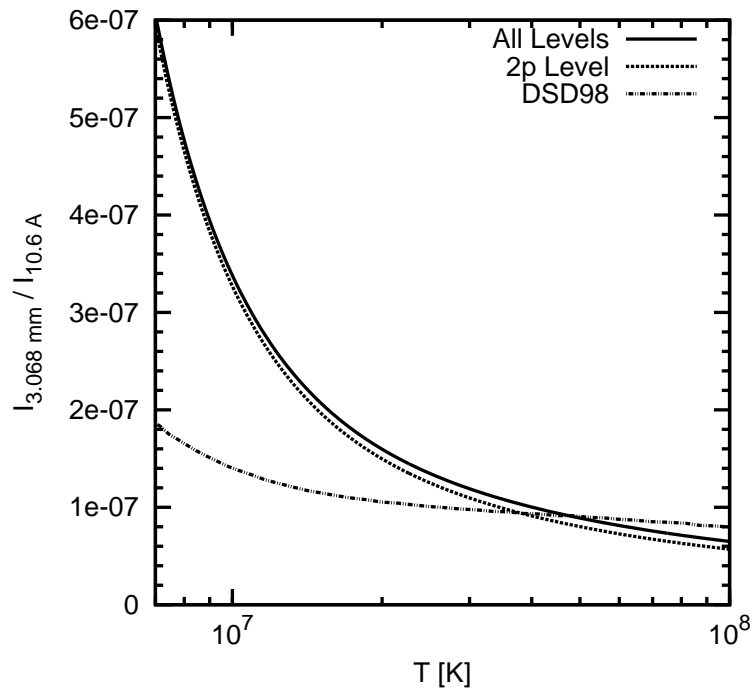}
		\caption{
			{\em (Left panel)}
			Effective collision strength as a function of temperature.
			{\em (Right panel)}
			Hyperfine line intensity in units of the 10.6\unit{\AA} X-ray
			doublet.
			In both plots, the excitations shown are direct (dot-dashed),
			indirect through the $2p$ level (dashed), and through all
			higher levels in \cloudy{} (solid).
			The results of \citet{DCruzSarazin1998} are also shown
			(dash-dot-dotted).
			\label{fig:iron-line-comp}
			}
	\end{centering}
\end{figure*}
\end{landscape}

The importance of these improvements for the lithium-like
\ironiso{} line is illustrated in Figure~\ref{fig:iron-line-comp}.
The collision strengths for direct, indirect through the $2p$ level,
and indirect excitations through all higher levels are compared to
the results of DSD98 in the left panel.
In agreement with these researchers, inclusion of indirect collisional
excitations from higher levels than $2p$ increases the effective collision
strength by a modest amount that approaches 15\% at 10$^8$~K.
The collision strength due to the $2p$ indirect collisions is higher than
the one reported in Figure~5 of DSD98, probably due to improvements in
atomic physics data since that publication. 
Overall, the updated collision strengths are higher than those reported
by DSD98 by 40--75\%.

\par
To complete the comparison with the DSD98 findings, the right panel of
Figure~\ref{fig:iron-line-comp} presents the hyperfine line intensity
in units of the 10.6\unit{\AA} X-ray doublet.
The improved atomic physics leads to a boost in the line ratio at lower
temperatures.
At the temperature of the ionization fraction peak ($\sim$20\unit{MK})
it is roughly 50\%.

\par
\begin{landscape}
\begin{figure*}
	\begin{centering}
		\includegraphics[scale=0.660]{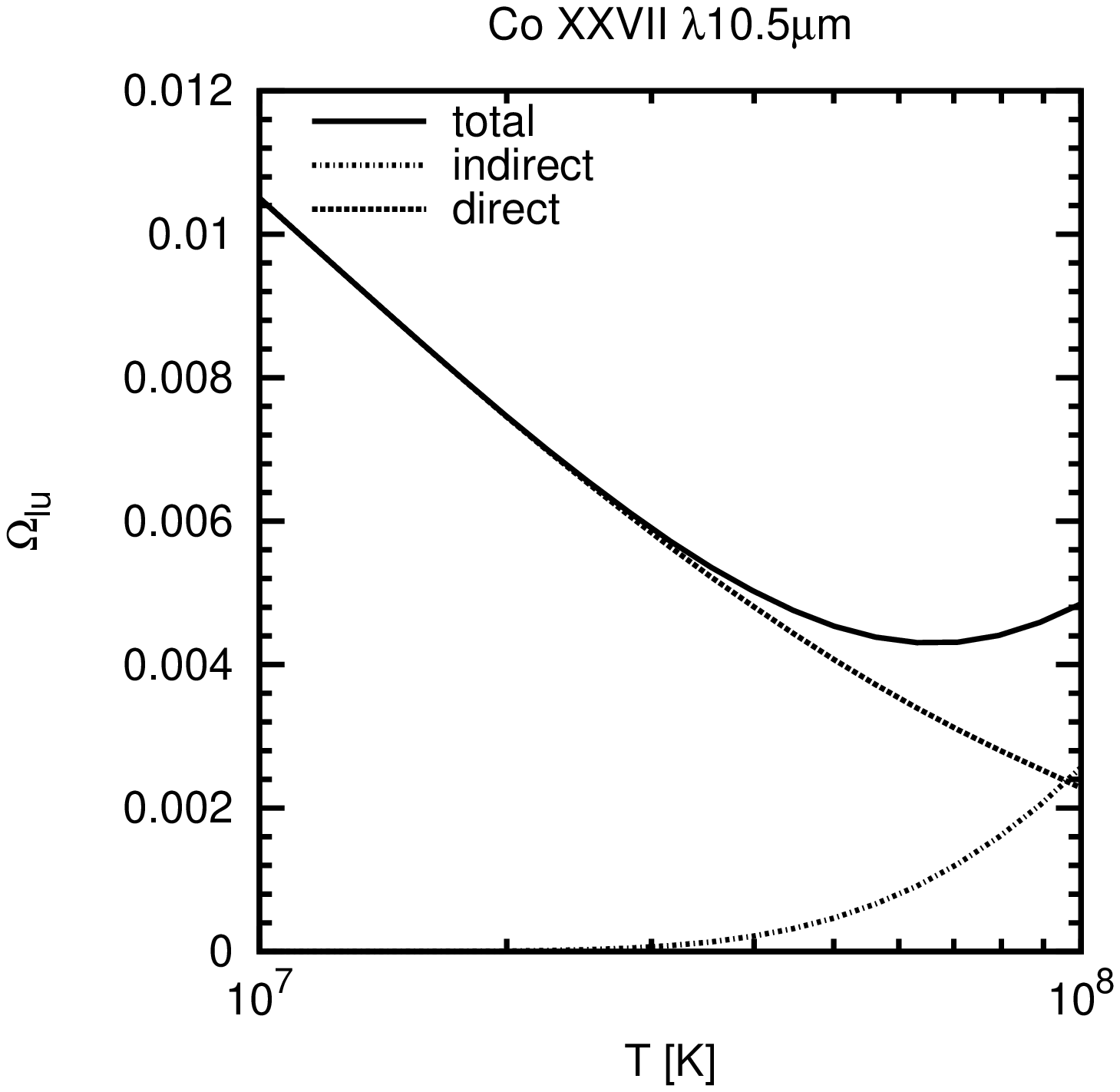}
		\includegraphics[scale=0.660]{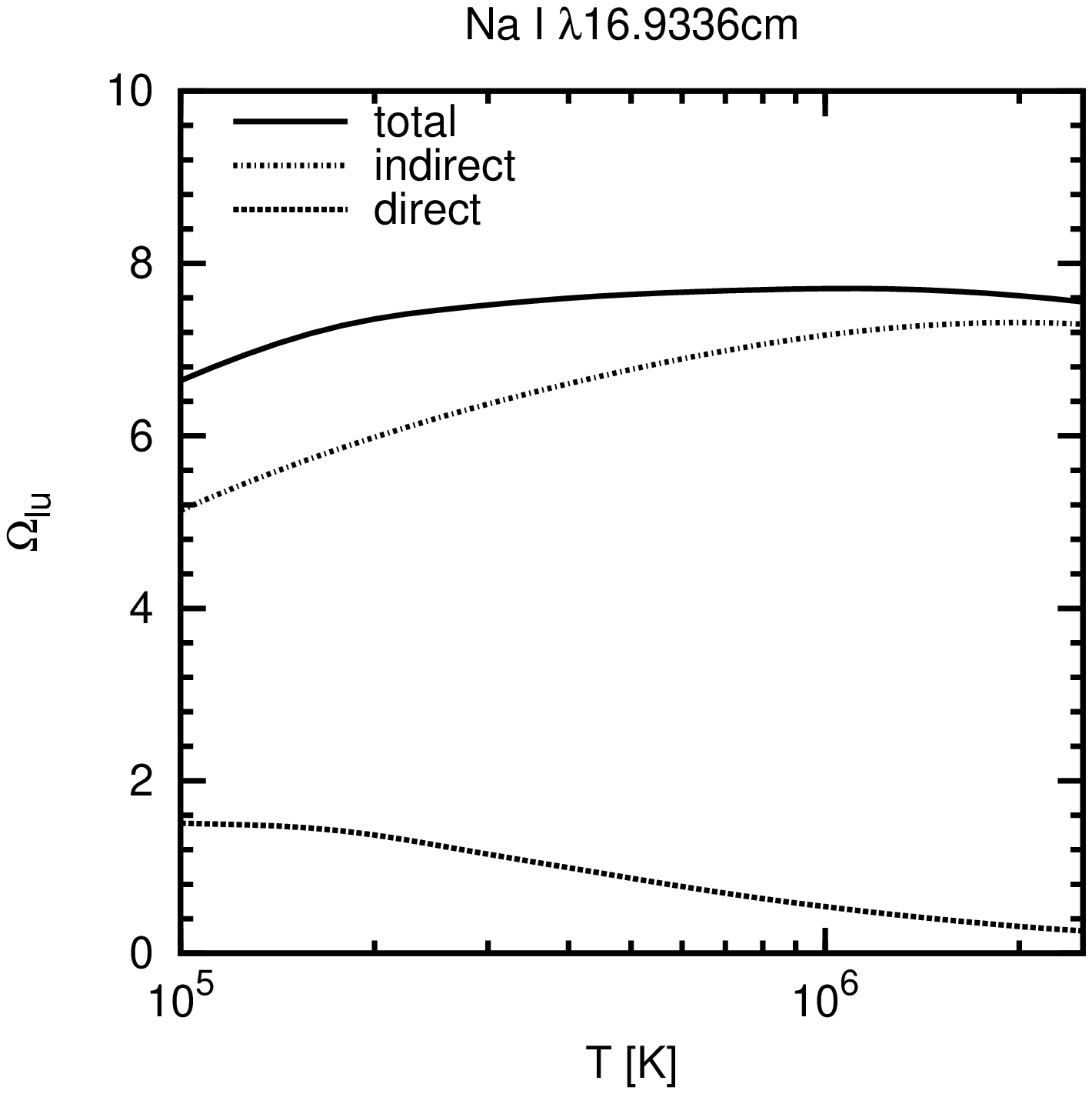}
		\caption{
			Collision strength decomposition for hydrogenic cobalt
			{\em (left panel)} and neutral sodium {\em (right panel)}.
			The dominant mode of collisional excitation depends on
			the iso-sequence.
			\label{fig:cs-hlike-nalike}
			}
	\end{centering}
\end{figure*}
\end{landscape}

The relative importance of direct and indirect excitations varies across
elements and isoelectronic sequences.
In the simplest case of hydrogenic ions, the energy above ground varies
as $Z^2$, so that indirect excitations are not particularly important
for heavy ions (e.g., Fig.~\ref{fig:cs-hlike-nalike}, left panel).
For the lithium-like and sodium-like sequences, the K and L shells are
respectively full, the energy separation between the ground and the first
excited state decreases, indirect excitations become favorable, and the
relevant collision strength attains large values
(e.g., Fig.~\ref{fig:cs-hlike-nalike}, right panel).


\subsection{Optical Pumping}\label{sec:optical-pumping}

\par
External radiation fields may affect line emissivity.
The Cosmic Microwave Background (CMB) plays an important
role for most of the lines considered here, and is included
in all the models below.
After correcting for the CMB continuum, line emissivity is
reduced by a factor (see DSD98, eqn.~(35), (36)) that depends
on the radiation temperature, and atomic physics.
This correction is implemented in \cloudy{} according to
\citet{Chatzikos-pumping}.

\par
On the other hand, radiation from a point source that lies away from
the sightline can lead to observable line emission enhancements.
This may be accomplished by direct or indirect radiative excitations
that populate the upper hyperfine level.
Although the CMB photon occupation number typically exceeds that due to
point sources (unless in close proximity to the source), the fact that
the radiation does not contribute to continuum translates to a
significant boost in the line emissivity.
We refer to this process as ``optical pumping''.

\par
In the case of galaxy clusters, the central Active Galactic Nucleus (AGN)
can affect the hyperfine line emissivity, and even the ionization structure
of the gas.
For illustration, we consider the radiation field of \ngc{} at the center of
the Perseus Cluster, although it is an unusually bright source.
The compilation of its Spectral Energy Distribution (SED) is presented in
Appendix~\ref{app:sec:seds}, where special care is taken to account for
frequency-dependent, long-term variability.

\par
The effect of the SED is best demonstrated by examining the properties of the
\ironXXIVline{} $\lambda$3.068\unit{mm} line as a function of source proximity.
The gas density is set to $10^{-3}\unit{\pcc}$ and the volume to 1\unit{\cc}.

\par
\begin{landscape}
\begin{figure*}
	\begin{centering}
		\includegraphics[scale=0.625]{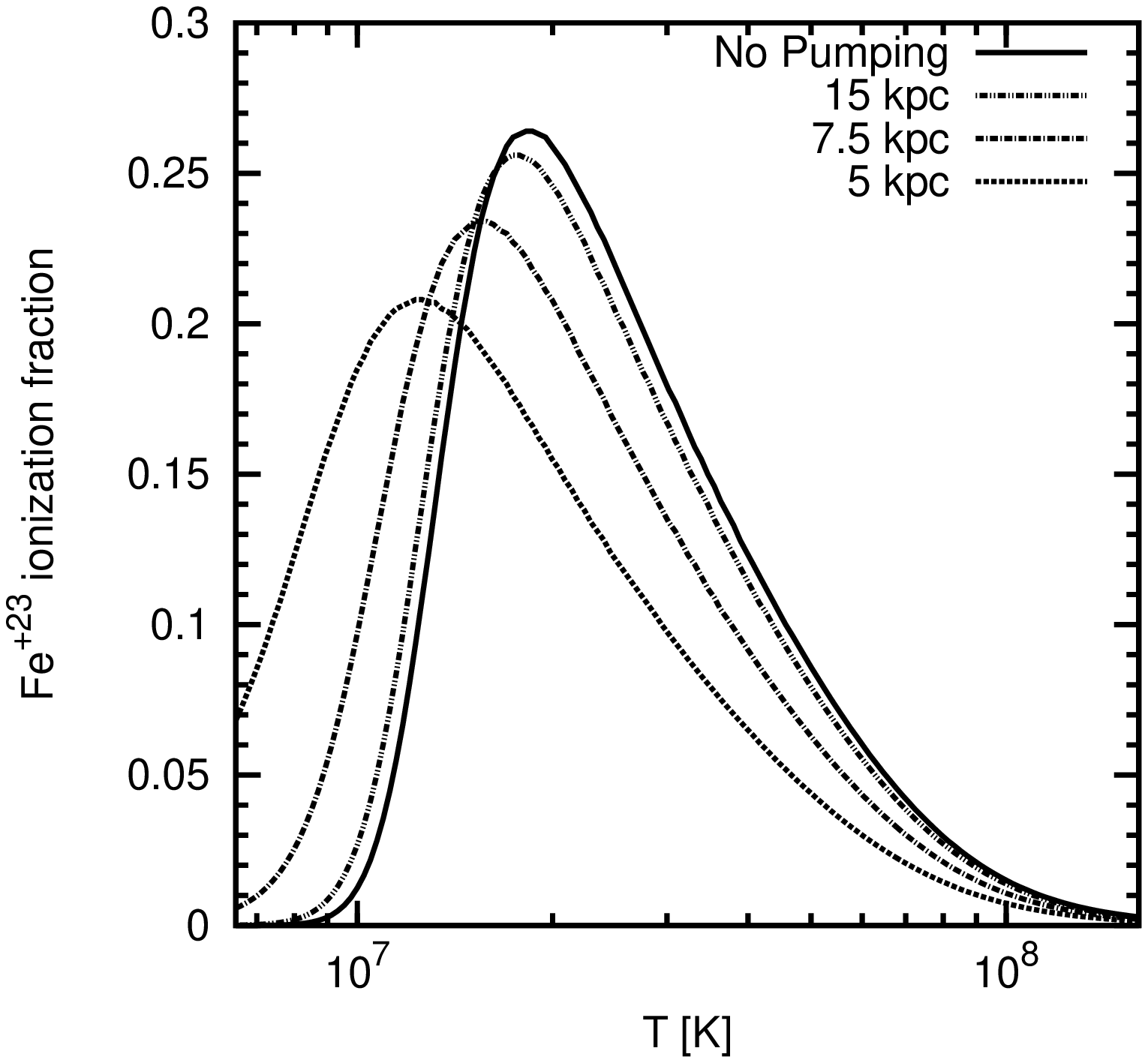}
		\includegraphics[scale=0.625]{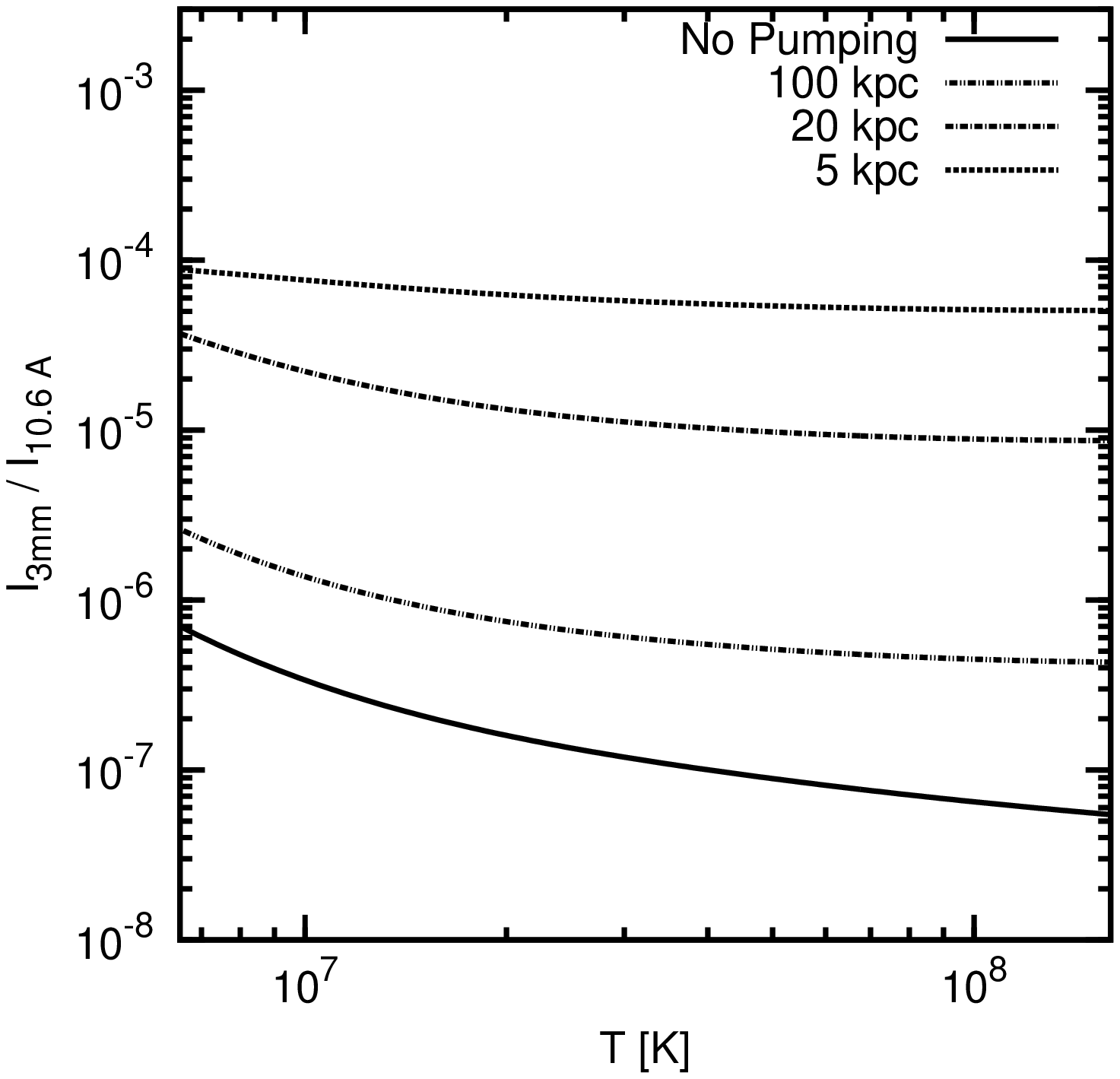}
		\caption{
			Temperature variation of
			{\em (left panel)} the ionization fraction of \ironXXIV{} and
			{\em (right panel)} the ratio of the hyperfine line
			over the 10.6\unit{\AA} X-ray doublet,
			as a function of point source proximity.
			Solid lines illustrate purely collisional results,
			while other line types represent results at various
			central distances, as indicated in the keys.
			\label{fig:sed-effects}
			}
	\end{centering}
\end{figure*}
\end{landscape}

The left panel in Figure~\ref{fig:sed-effects} shows the ionization
fraction of \ironXXIV{} as a function of temperature and distance to
the point source.
The SED acts to decrease the maximum ionization fraction for this species,
and to shift that maximum to lower temperatures.
At 5\unit{kpc} the maximum fraction and its temperature are reduced by factors
of 20\% and 2, respectively, relative to the purely collisional case.
Because the number of ionizing photons decreases as $1/r^2$, $r$ being the
distance to the source, these effects are quickly moderated with increasing
distance and they are minimal by 15\unit{kpc}.

\par
The right panel in Figure~\ref{fig:sed-effects} shows the line emissivity as a
function of source proximity.
The emissivity at 5\unit{kpc} is at least two orders of magnitude higher than
the purely collisional case at all temperatures.
The boost is about 2 orders of magnitude at 20\unit{kpc}, and remains within a
factor of a few even at 100\unit{kpc}.

\par
Note that less bright sources lead to more modest effects (Section \ref{sec:clusters}).

\section{Intensity Grids}\label{sec:intensity-maps}

\par
Because the vast majority of the hyperfine lines are optically thin for a wide
range of conditions, it is beneficial to compile intensity grids as a function
of the gas density and temperature.
Such calculations are presented in the following.

\par
At each grid point, continuum-corrected line intensities are computed for a unit
volume of gas of definite density and temperature, exposed to the CMB.
The temperature varies in the range $10^4$--$10^8\unit{K}$, in which the relevant
ions are expected to be most abundant, while the density varies in the range
$10^{-3}$--$10^{+3}$\unit{\pcc}.

\par
\begin{landscape}
\begin{figure*}
	\begin{centering}
		\includegraphics[scale=0.41]{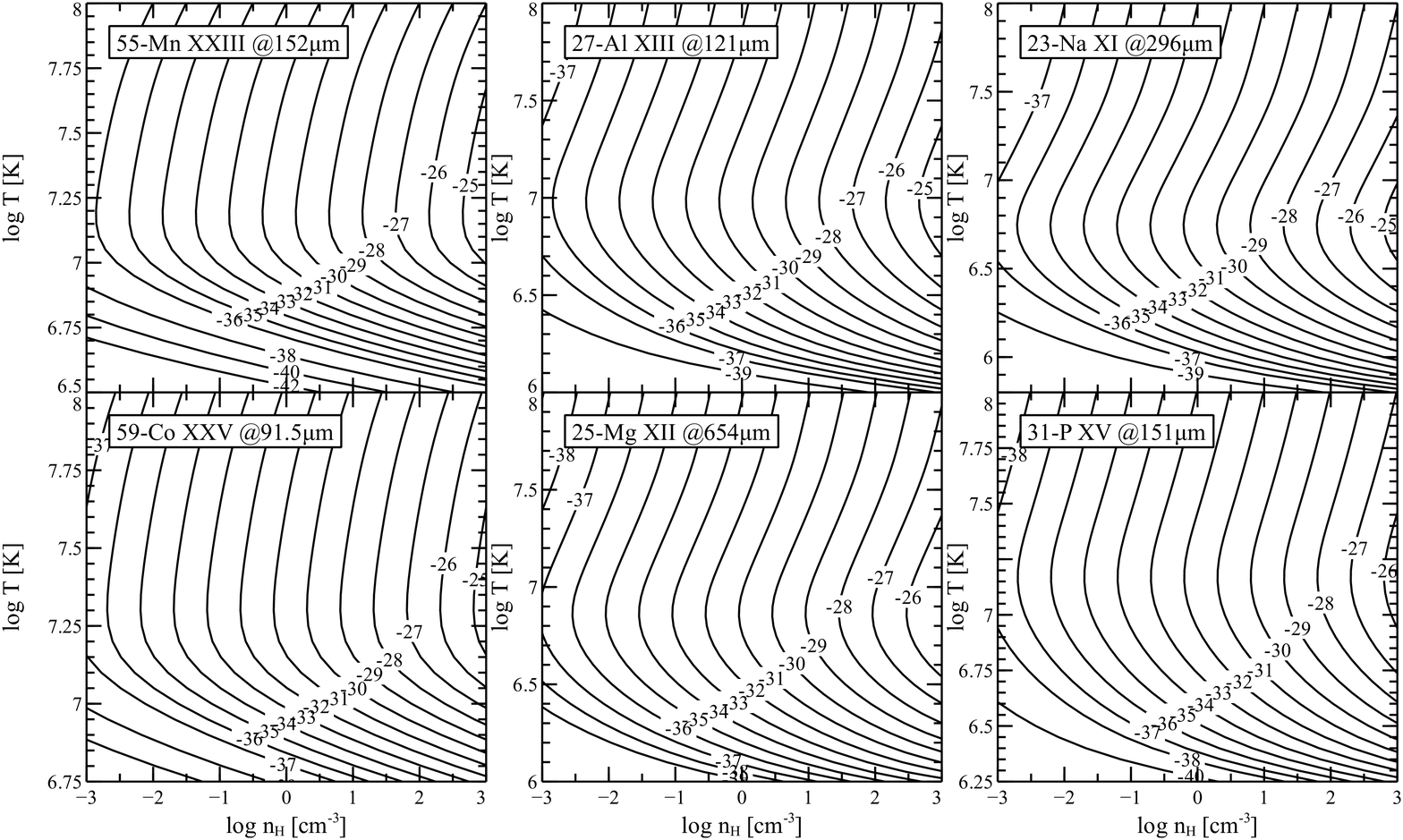}
		\caption{
			Log intensity maps for the coronal hyperfine lines
			in order of descending peak intensity from left
			to right, and from top to bottom.
			Figures \ref{fig:coronal-lines}.1--\ref{fig:coronal-lines}.14
			are available in the online version of the Journal.
			\label{fig:coronal-lines}
			}
	\end{centering}
\end{figure*}
\end{landscape}

Figure \ref{fig:coronal-lines} presents the intensity contour maps.
The quantity plotted in the contour maps below is the intensity integrated over
all angles and the line profile, and has units of log flux.
The maps are sorted in order of descending peak intensity.

\par
All maps follow a similar outline.
At a given temperature, the intensity increases with the density.
The critical density of the two-level system may be defined as
\begin{equation}
	n_\mathrm{e,c} = \frac{g_l \, A_\mathrm{ul} \, T^{1/2}}{8.629 \times 10^{-6} \, \Omega^\mathrm{eff}} \, (1 + \eta_\nu) \,
	\label{eqn:ncrit}
\end{equation}
where $g_l$ is the statistical weight of the lower hyperfine level, and
$\eta_\nu$ is the CMB photon occupation number at the line frequency.
Obviously, the exact value of the critical density for a given line
depends on the transition probability, as well as the effective collision
strength.
Sodium-like lines tend to have critical densities below our lowest limit,
and vary as the first power of the density.
On the other hand, hydrogenic lines tend to have critical densities above
our highest limit, and vary as the square of the density.
Finally, the critical densities for most lithium-like lines fall within
the covered range and vary as $n^2$ at low densities, and as $n$ at high
densities.

\par
On the other hand, at a given density, the intensity rises quickly with
temperature, reaches a peak at some characteristic value at which the
ionization fraction is maximum, and then declines.
For hydrogenic atoms, the line intensity contours at higher temperatures
are convex (positive second derivative), while for lithium-like and sodium-like
atoms the contours are concave and, in some cases, bend sharply to higher densities.
In addition, the Li-like and Na-like contours of second- and third-row elements
possess a secondary intensity peak at higher temperatures.

\par
These types of behavior at high temperatures may be understood as follows.
Firstly, the slow variation of the contours of hydrogenic atoms arises
from the balance between collisional ionization and radiative recombination
between the hydrogenic ions and the corresponding bare nuclei.
In this situation, the electron density drops out of the balance
equation, and the density ratio of these ionization stages is equal
to the ratio of the recombination coefficient over the collisional
ionization coefficient, which varies very slowly with temperature.

\par
On the other hand, in lithium-like and sodium-like ions, such a balance cannot
be achieved because the ion can be further ionized.
The ion abundance decreases quickly with temperature, at a rate regulated by the
ionization potential.
As a result, the intensity contours for lithium-like ions of fourth-row elements
decrease gently, while for second- and third-row elements, as well as for all
sodium-like ions, the contours decrease more sharply.

\par
As the temperature increases, the closed shell is progressively ionized.
Some of the released electrons may recombine with lower stage ions, e.g.,
electrons released from He-like carbon may recombine with Li-like carbon.
The second intensity peak seen in lithium-like and sodium-like ions of second-
and third-row elements arises due to the fact that dielectronic recombination
with these electrons is more important than radiative recombination.
However, the deeper potential wells of fourth-row elements disfavor
the excitations necessary for dielectronic recombination, and moderate
its importance.

\subsubsection{Normalizing to the 20\AA{} X-ray Bump}

\par
As an observational aid, we express the intensity of the coronal lines
relative to the intensity of the line complex at 20\unit{\AA} (18--28\unit{\AA},
or 0.45-0.70\unit{keV}), including the bremsstrahlung continuum.

\par
\begin{landscape}
\begin{figure*}
	\begin{centering}
		\includegraphics[scale=0.40]{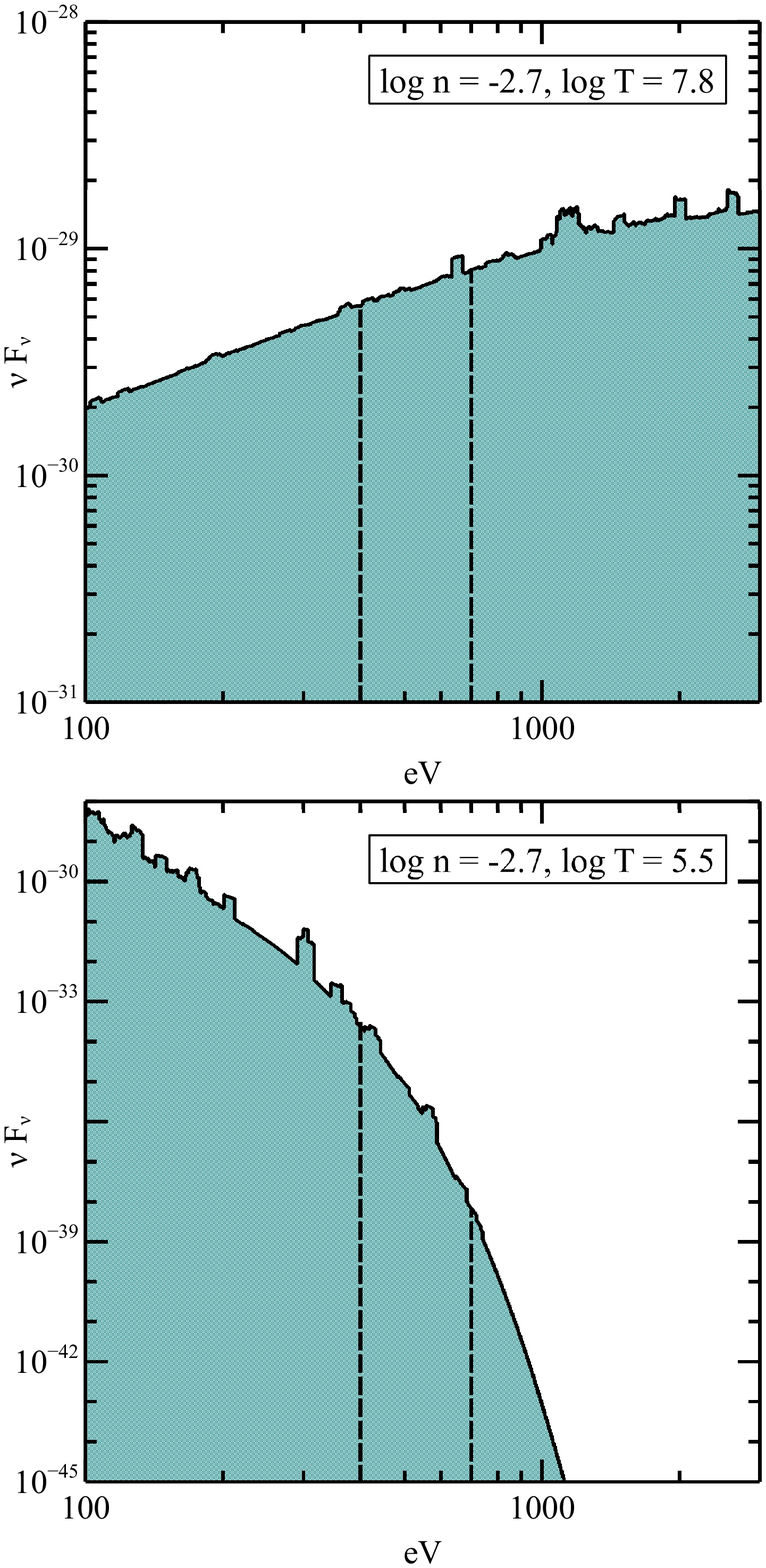}
		\includegraphics[scale=0.55]{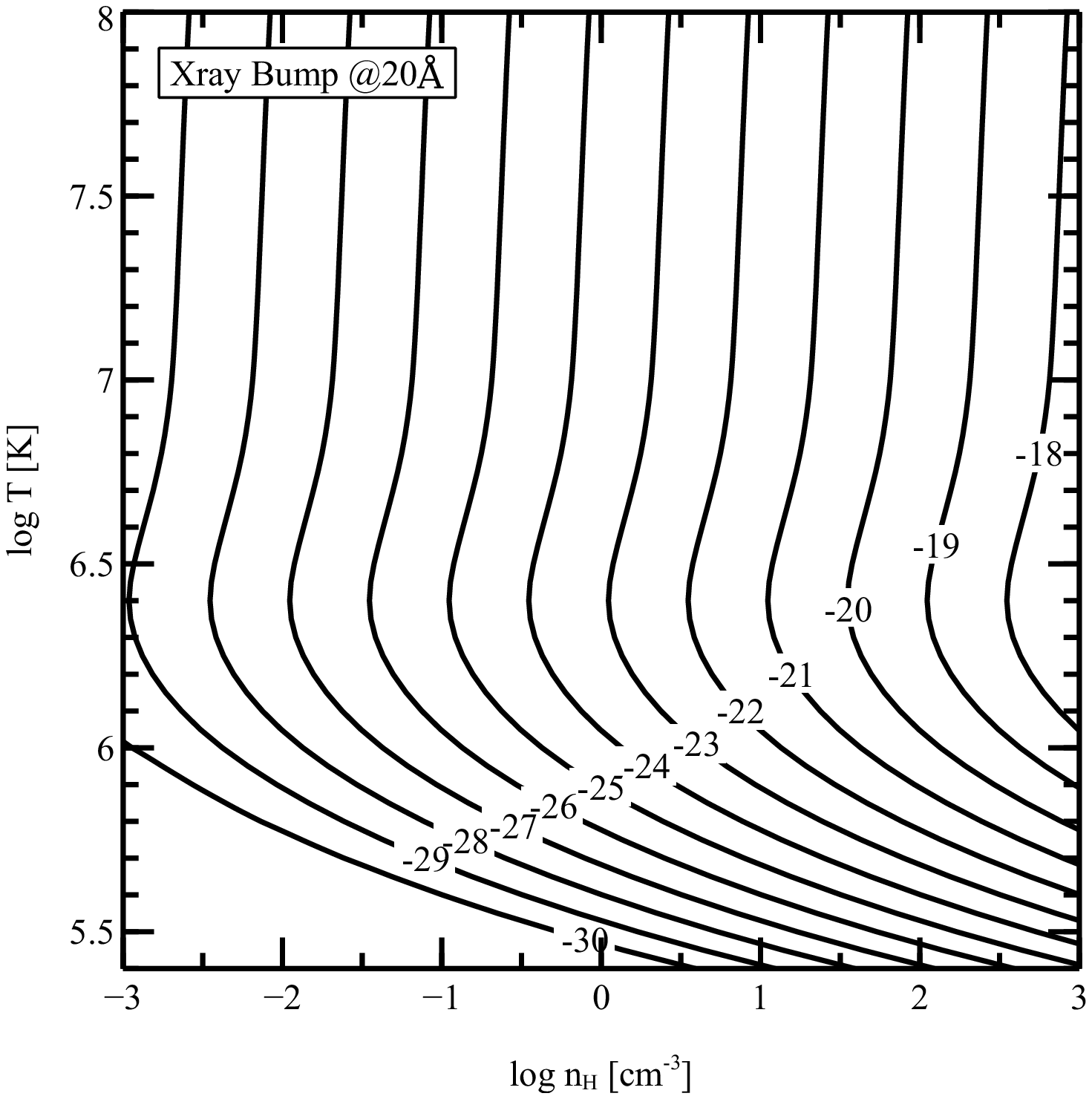}
		\includegraphics[scale=0.40]{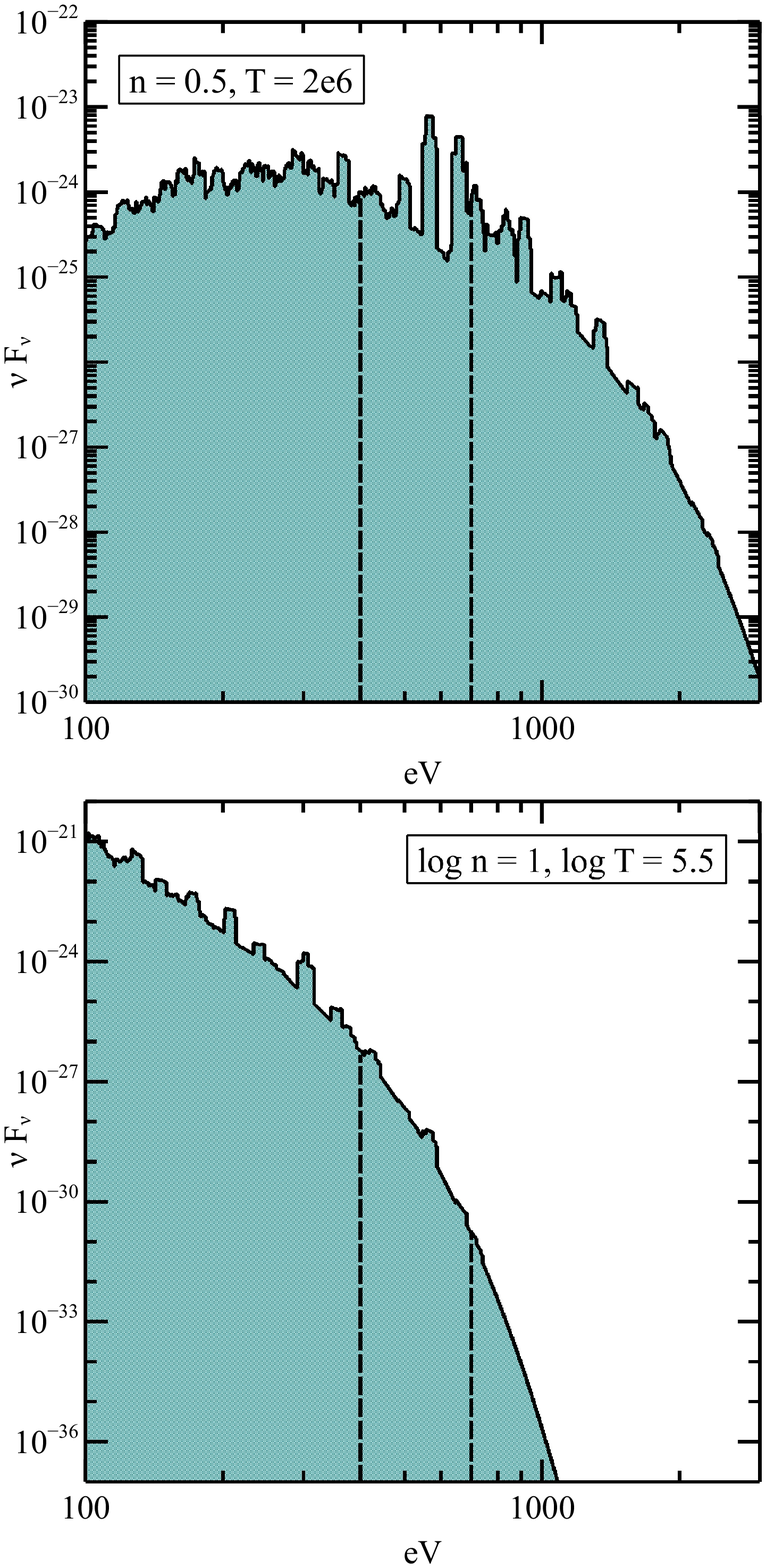}
		\caption{
			{\em (Center)}
			Intensity contours of the X-ray complex at 20\unit{\AA}
			(see text), including the bremsstrahlung continuum.
			{\em (Left \& Right)}
			X-ray spectra in the range 0.1--3\unit{keV} for various
			points in the grid, as indicated in the legends.
			The density and temperature for the top right spectrum
			are similar to the conditions of the hot gas in the
			Extended Orion Nebula \citep{Orion-XMM}.
			The spectra are at higher resolution than typically
			attainable in X-ray observations.
			\label{fig:xray-complex}
			}
	\end{centering}
\end{figure*}
\end{landscape}

For reference, the intensity contours of the X-ray complex
are shown in Figure~\ref{fig:xray-complex}, along with the
spectra of various physical conditions of interest.
At high temperatures the contours are nearly vertical
due to the fact that the intensity for both the continuum
and the line emission varies with the square of the density,
with a slow dependence on temperature.

\par
\begin{landscape}
\begin{figure*}
	\begin{centering}
		\includegraphics[scale=0.41]{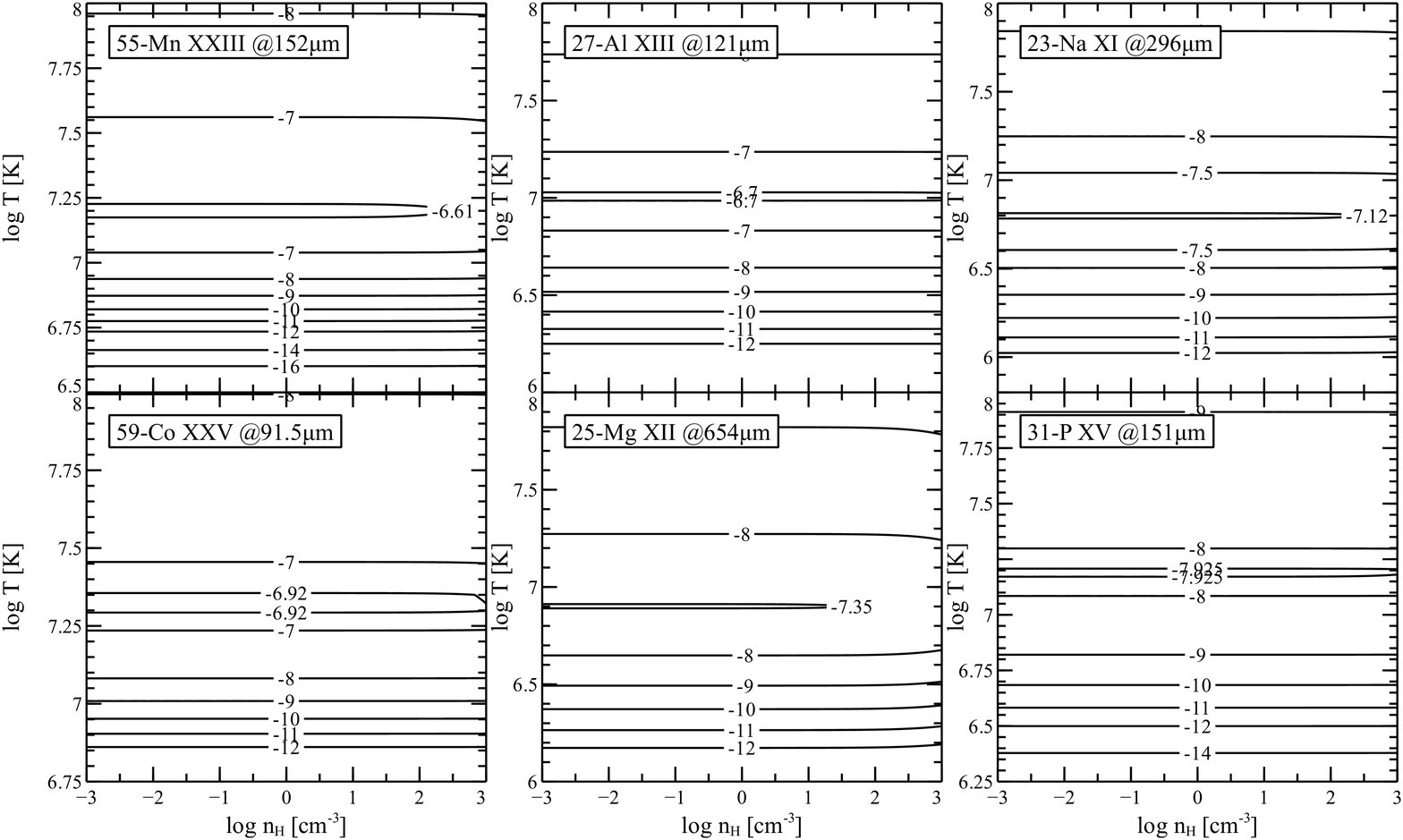}
		\caption{
			Intensity maps for the coronal hyperfine lines normalized
			by the intensity of the X-ray line complex.
			The images are presented in the same order as Figure~\ref{fig:coronal-lines}.
			Figures \ref{fig:coronal-lines-norm}.1--\ref{fig:coronal-lines-norm}.11
			are available in the online version of the Journal.
			\label{fig:coronal-lines-norm}
			}
	\end{centering}
\end{figure*}
\end{landscape}

Figure~\ref{fig:coronal-lines-norm} shows the contours of the normalized
line flux, for those ions whose ionization fraction obtains its highest
value at temperatures above 10$^{5.4}$\unit{K}.
Below the critical density the contours are have only a temperature
dependence.
More complicated contour outlines are manifest above the critical density.
Accordingly, most hydrogenic lines and lithium-like lines of heavy elements
have density-independent contours, contrary to lines that arise from shallower
ionization potentials.


\section{Hyperfine Line Predictions}\label{sec:predictions}

\par
In this section we use \cloudy{} to compute the hyperfine line intensities
that arise from hot gas in the Perseus and Virgo galaxy clusters, as well
as in the Orion Nebula.

\par
\cloudy{} reports the emissivity integrated over the line profile and over all angles.
We convert this to the spectral emissivity at the line center by using the formula
\begin{equation}
	j_\nu = \frac{j}{4 \pi^{3/2} \Delta \nu_D}	\,	,
	\label{eqn:j-spec}
\end{equation}
where $\Delta \nu_D$ is the Doppler line width.
The $4 \pi$ term accounts for the angular dependence, while the remainder is
the profile at the line center, that is,
$\phi(\nu_0) = \{ \sqrt{\pi} \, \Delta \nu_D \}^{-1}$.

\par
We deduce the profile at the line center as follows.
We assume that the line follows the Voigt profile, balanced by the competition
between natural (radiative and pressure) broadening on one hand, and Doppler
broadening on the other.
This is encapsulated by the parameter
$\alpha = (\gamma^\mathrm{rad} + \gamma^\mathrm{col}) / (4\pi \, \Delta \nu_D)$,
where the $\gamma$'s are the rates of radiative ($\gamma^\mathrm{rad} = A_\mathrm{ul}$)
and collisional damping.
Because pressure broadening at the line core is best described by the impact
approximation, we take the damping rate equal to the collisional deexcitation
rate, $\gamma^\mathrm{col} = C_\mathrm{ul}$.
The parameter, then, obtains vanishingly small values ($\alpha \ll 1$),
and the (unnormalized) Voigt function at the line center is
roughly 1, leading to the equation above.

\par
Note that in the following the Doppler width does not include a turbulent
component, so our intensity estimates in the context of galaxy clusters
should be treated as upper limits.

\subsection{Galaxy Clusters}\label{sec:clusters}

\par
Realistic line predictions require accurate descriptions of the
distribution of the ICM density, temperature, and metallicity
with distance from the cluster center.
For that purpose, we employ \chandra, \xmm, and \rosat{}
observations as described in detail below.

\par
For all hyperfine lines, we compute the (CMB-corrected) emissivity
radial profiles.
All lines are optically thin when integrated through the
cluster diameters, with the highest optical depth in
Perseus having a value of about $10^{-4}$.
The transfer of the hyperfine lines can then be ignored, and
intensity profiles can be computed by simply integrating the
emissivity profiles along the line of sight.

\par
In the tables below, for each line we present three estimates for
the intensity: at the peak of the intensity profile; for a 0.1
square arc-minute fiducial aperture centered on the cluster center;
and for the entire cluster volume.
Brightness temperatures are provided for lines with
$\lambda > 1\unit{mm}$ as well.
The line flux is also expressed in units of the X-ray
continuum flux.

\subsubsection{Perseus Galaxy Cluster}\label{sec:perseus}


\par
We model the ICM in Perseus with the aid of the \xmm{} observations
of \citet{Churazov-XMM-Perseus} for the plasma within the central
200\unit{kpc}, and of the \rosat{} observations of
\citet{Ettori-ROSAT-Perseus} for the plasma in the cluster outskirts.
For the metallicity in the outskirts we use the average metallicity
of the out-most bins of the \xmm{} observation (0.48 solar).
Note that the substructure at the cluster center revealed with \chandra{}
\citep{Fabian-Perseus-2006} is not expected to alter our conclusions
significantly, and has not been accounted for in these calculations.

\par
\begin{landscape}
\begin{figure*}
	\begin{centering}
		\includegraphics[scale=0.625]{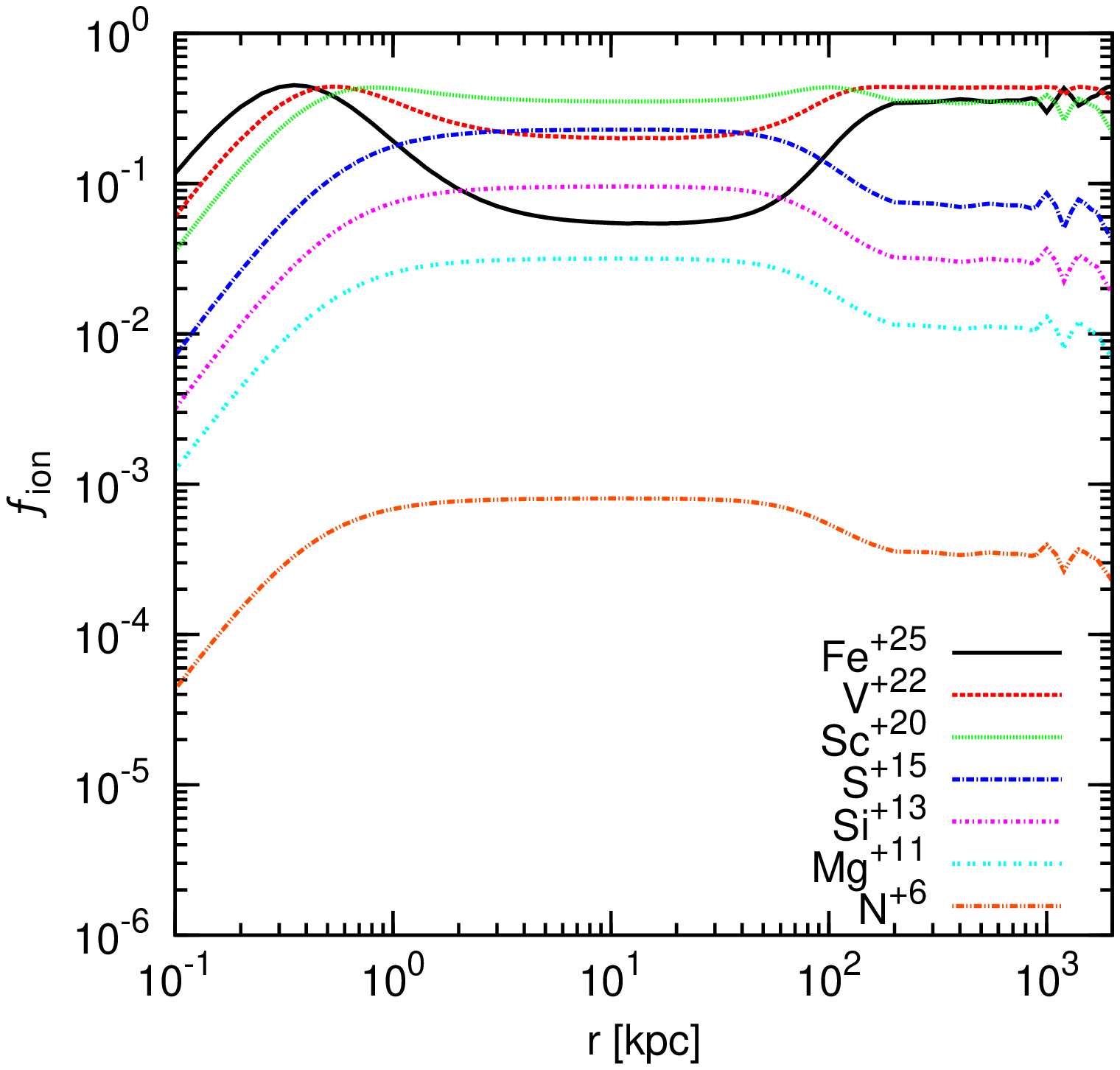}
		\includegraphics[scale=0.625]{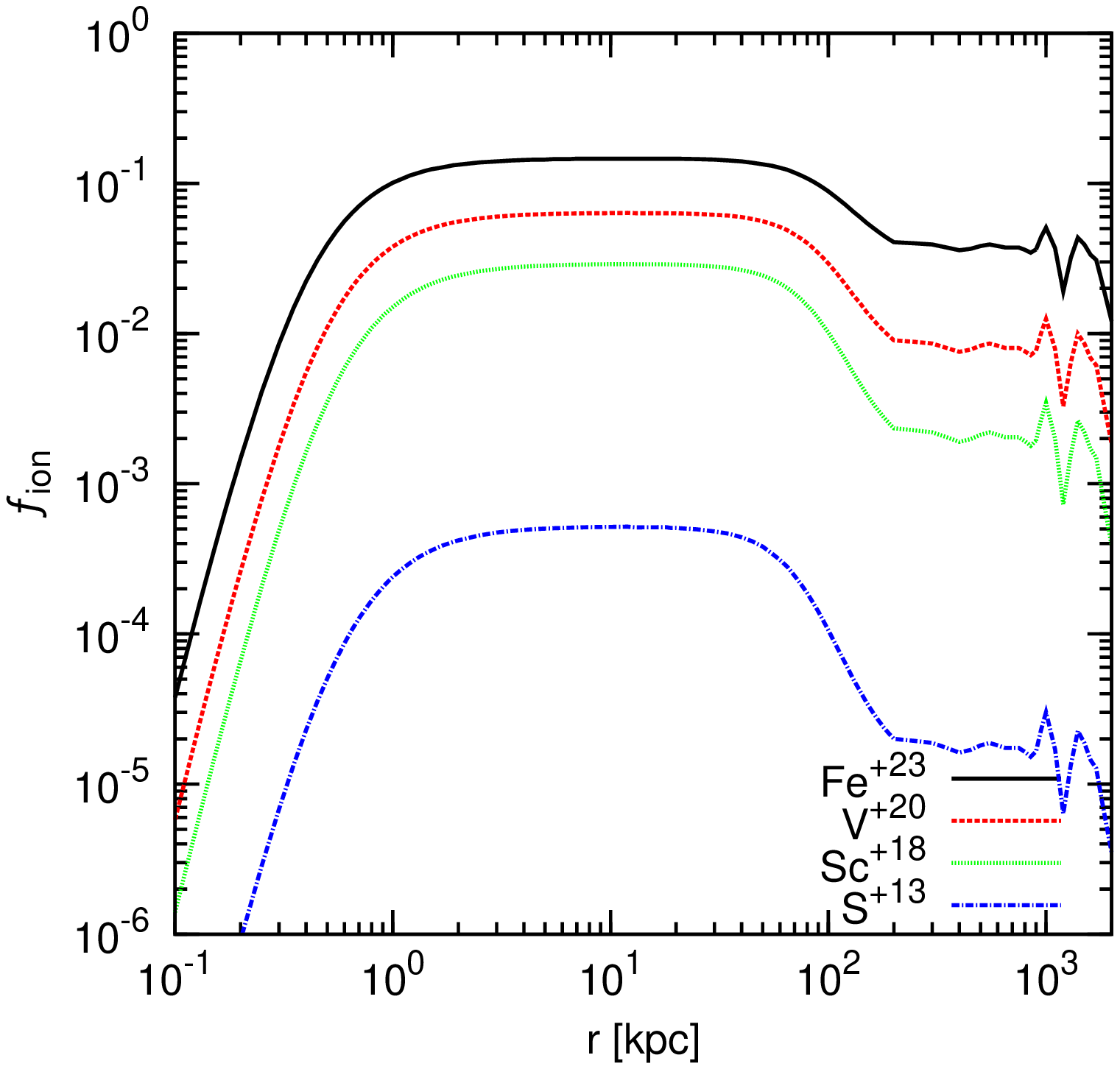}
		\caption{
			Example ionization fraction profiles for hydrogenic {\em (left panel)}
			and lithium-like {\em (right panel)} ions.
			The effects of the AGN radiation are prominent within the central
			2\unit{kpc}.
			Note that the temperature profile rises beyond 100\unit{kpc}
			from a central floor of about 3\unit{keV} to about 6.5\unit{keV}
			at 900\unit{kpc}.
			The structure seen around 1\unit{Mpc} is due to uncertainties
			in the input \rosat{} temperature profile.
			\label{fig:perseus-ion-fractions}
			}
	\end{centering}
\end{figure*}
\end{landscape}

As suggested in Section \ref{sec:optical-pumping}, the presence of the AGN
can greatly affect the ionization structure of the ICM near the cluster
center.
Figure~\ref{fig:perseus-ion-fractions} illustrates the effects of the
intense radiation on the ionization structure of the plasma.
The ions shown were selected because they are relevant to X-ray
spectroscopy obtained from cool core clusters \citep{Peterson-CoolCore-Spectra}.
In the absence of the ionizing background, the ion fractions
are constant within the temperature core of the cluster
($r \lesssim 50\unit{kpc}$).
By contrast, the AGN radiation field significantly alters
the ionization structure of the gas within the central 2\unit{kpc},
reducing the fractional densities of species with shallow ionization
potentials and enhancing those with deep potentials.
In our calculations, the densities of hydrogenic ions of elements
with $21 \leq Z \leq 27$ are increased by up to 1 order of magnitude.

\par
Because the effect is localized to the vicinity of the AGN,
the altered ionization fractions do not affect our estimates
for (projected) line intensities much.
However, the AGN may significantly impact the ionization structure
of the interstellar medium near the center of the brightest cluster
galaxy \ngc{}.
A detailed exploration of this possibility is beyond the scope of
this paper.

\par
\begin{landscape}
\begin{figure*}
	\begin{centering}
		\includegraphics[scale=0.625]{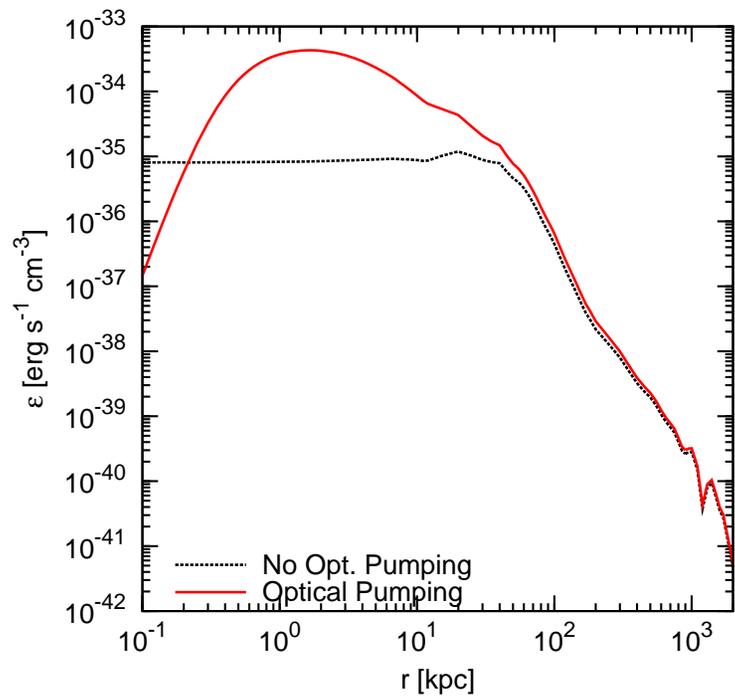}
		\includegraphics[scale=0.625]{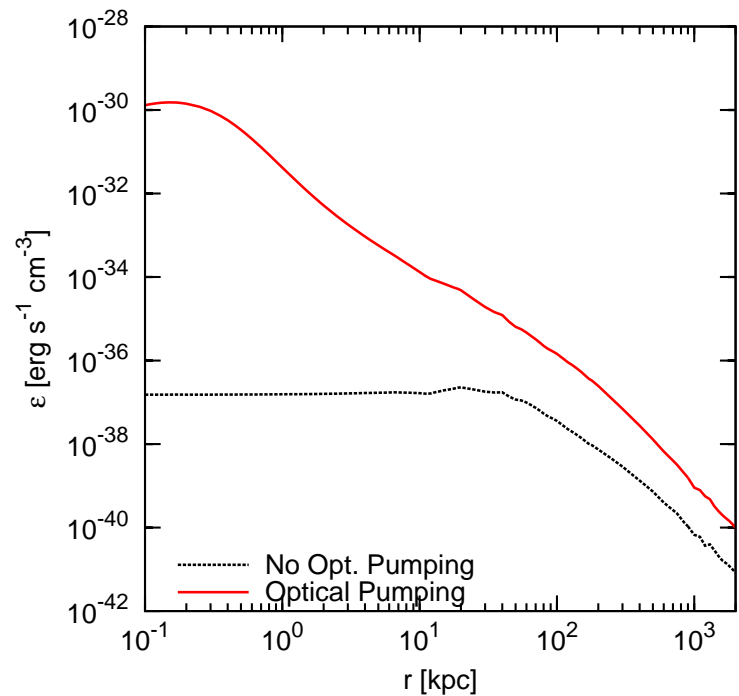}
		\caption{
			Comparison of the predicted emissivity profile of the
			$^{57}$\ion{Fe}{24} $\lambda$3.068\unit{mm} {\em (left panel)}, and
			the $^{57}$\ion{Fe}{26}  $\lambda$348\unit{\mu m} {\em (right panel)}
			lines in Perseus, with the effects of optical pumping accounted for
			and ignored.
			\label{fig:perseus-iron-line}
			}
	\end{centering}
\end{figure*}
\end{landscape}

Line emissivities are affected by the AGN through the ionization fractions,
as well as through optical pumping.
Figure~\ref{fig:perseus-iron-line} illustrates that for the physical conditions
prevalent at the cluster core, the emissivity of the \ironXXIVline{} line reaches
a boost of $\sim 50$ within the central 2\unit{kpc}, but plummets near the very
center, due to the modified ion densities.
On the other hand, the $^{57}$\ion{Fe}{26} line is boosted by about 7 orders of
magnitude near the AGN, by about 3 orders of magnitude at 10\unit{kpc}, and by
about 1 order of magnitude even in the cluster outskirts.
Clearly, optical pumping by the central AGN can affect the emissivities of certain
lines throughout the cluster volume.

\par
Table~\ref{table:perseus-bright-lines} lists the brightest lines in Perseus
in order of decreasing average intensity within the aperture.
The lines are in the mm or sub-mm part of the spectrum, and they are all
enhanced by optical pumping, similarly to the iron lines.

\par
\begin{figure*}
	\begin{centering}
		\includegraphics[scale=0.8]{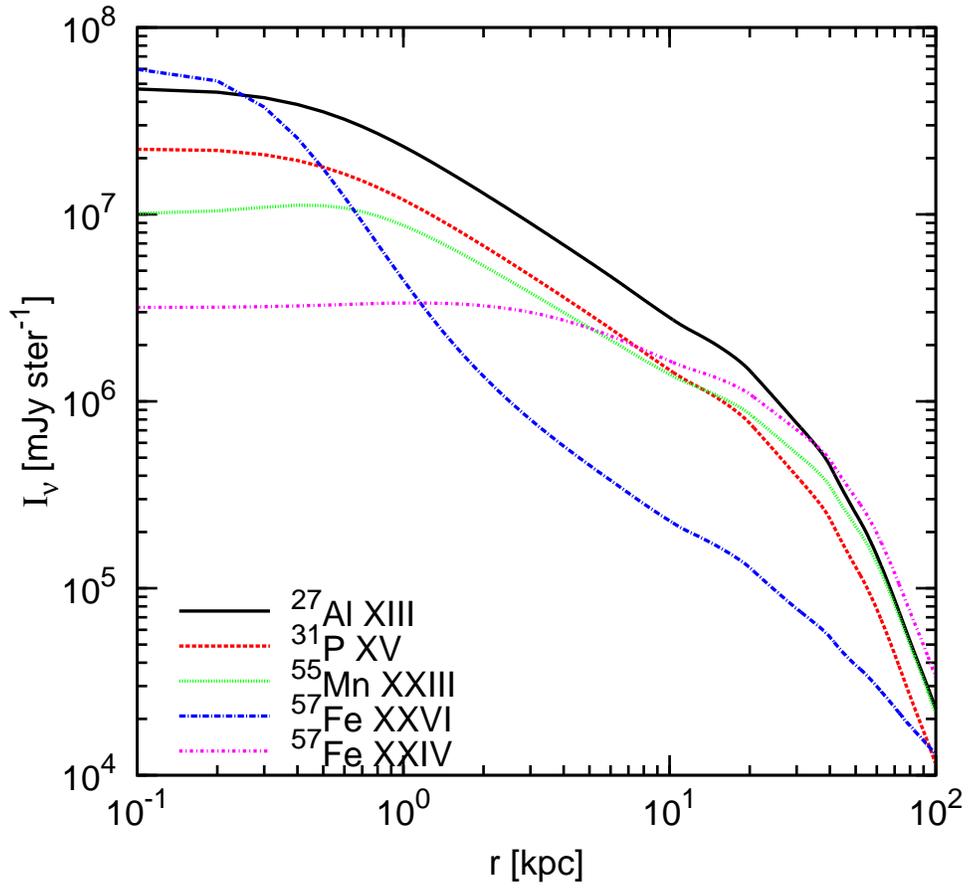}
		\caption{
			Intensity profiles within (projected) 100\unit{kpc}
			of \ngc{} for the five brightest hyperfine lines.
			\label{fig:perseus-line-intensity}
			}
	\end{centering}
\end{figure*}

Figure~\ref{fig:perseus-line-intensity} presents the intensity profiles
for the five brightest lines in Perseus, that is, their emissivities
integrated along the line of sight as a function of the projected distance
from the cluster center.
The diversity in profile shapes is caused by the intense radiation field of
the AGN.


\begin{landscape}
\begin{deluxetable}{lcccccccc}
\tabletypesize{\rm\small}
\tablecolumns{9}
\tablewidth{0pt}
\tablecaption{Perseus Brightest Lines\label{table:perseus-bright-lines}}
\tablehead
{
    Species                     &
    $\lambda$                   &
    I$_\nu^\mathrm{max}$        &
    T$_b^\mathrm{max}$          &
    I$_\nu^\mathrm{apr}$        &
    T$_b^\mathrm{apr}$          &
    I$_\nu^\mathrm{tot}$        &
    T$_b^\mathrm{tot}$          &
    F / F$_\mathrm{cont}$    \\
                                &
    (cm)                        &
    (mJy ster$^{-1}$)           &
    (mK)                        &
    (mJy ster$^{-1}$)           &
    (mK)                        &
    (mJy ster$^{-1}$)           &
    (mK)                        &
                             \\
    (1) &
    (2) &
    (3) &
    (4) &
    (5) &
    (6) &
    (7) &
    (8) &
    (9)
}
\startdata

$^{27}$\ion{Al}{13}  	& 1.210E$-$02	& 2.646E+07	&        ---	& 5.223E+06	&        ---	& 4.471E+02	&        ---	& 8.268E$-$07	\\
$^{31}$\ion{P}{15}   	& 1.510E$-$02	& 1.259E+07	&        ---	& 2.701E+06	&        ---	& 2.305E+02	&        ---	& 3.197E$-$07	\\
$^{55}$\ion{Mn}{23}  	& 1.520E$-$02	& 6.317E+06	&        ---	& 2.017E+06	&        ---	& 3.226E+02	&        ---	& 1.783E$-$07	\\
$^{57}$\ion{Fe}{26}  	& 3.480E$-$02	& 3.372E+07	&        ---	& 1.219E+06	&        ---	& 1.261E+02	&        ---	& 4.619E$-$08	\\
$^{57}$\ion{Fe}{24}  	& 3.068E$-$01	& 1.896E+06	& 6.4532E$-$03	& 1.197E+06	& 4.0747E$-$03	& 4.533E+02	& 1.5431E$-$06	& 5.167E$-$09	\\
$^{59}$\ion{Co}{25}  	& 9.150E$-$03	& 2.388E+06	&        ---	& 6.887E+05	&        ---	& 1.142E+02	&        ---	& 9.763E$-$08	\\
$^{25}$\ion{Mg}{12}  	& 6.540E$-$02	& 2.689E+06	&        ---	& 6.160E+05	&        ---	& 6.212E+01	&        ---	& 1.875E$-$08	\\
$^{35}$\ion{Cl}{17}  	& 2.130E$-$02	& 2.459E+06	&        ---	& 5.079E+05	&        ---	& 4.624E+01	&        ---	& 4.011E$-$08	\\
$^{39}$\ion{K}{19}   	& 3.170E$-$02	& 2.551E+06	&        ---	& 4.601E+05	&        ---	& 4.514E+01	&        ---	& 2.313E$-$08	\\
$^{29}$\ion{Si}{14}  	& 3.790E$-$02	& 1.987E+06	&        ---	& 4.172E+05	&        ---	& 4.029E+01	&        ---	& 2.035E$-$08	\\
$^{53}$\ion{Cr}{24}  	& 1.270E$-$02	& 6.458E+06	&        ---	& 3.461E+05	&        ---	& 3.628E+01	&        ---	& 3.726E$-$08	\\
$^{23}$\ion{Na}{11}  	& 2.960E$-$02	& 1.250E+06	&        ---	& 2.396E+05	&        ---	& 2.527E+01	&        ---	& 1.680E$-$08	\\
$^{33}$\ion{S}{16}   	& 3.270E$-$02	& 9.451E+05	&        ---	& 2.044E+05	&        ---	& 1.939E+01	&        ---	& 1.083E$-$08	\\
$^{14}$\ion{N}{7}    	& 5.652E$-$01	& 2.535E+05	& 2.9289E$-$03	& 1.675E+05	& 1.9355E$-$03	& 6.317E+01	& 7.2983E$-$07	& 7.919E$-$10	\\
$^{37}$\ion{Cl}{17}  	& 2.550E$-$02	& 6.552E+05	&        ---	& 1.368E+05	&        ---	& 1.267E+01	&        ---	& 8.778E$-$09	\\
$^{49}$\ion{Ti}{22}  	& 8.350E$-$03	& 1.157E+06	&        ---	& 1.065E+05	&        ---	& 1.089E+01	&        ---	& 1.814E$-$08	\\

\enddata
\end{deluxetable}
\end{landscape}

\subsubsection{Virgo Galaxy Cluster}\label{sec:virgo}

\par
We utilize the density and temperature fits of \citet{Churazov-Virgo-2008}
to model the ICM in the Virgo cluster.
We also make use of the detailed fits of \citet{Million-Virgo-Metals}
to describe the elemental abundance profiles of several elements
(Ne, Mg, Si, S, Ar, Ca, Fe, and Ni), and employ an average of these
profiles for other metals.

\par
Because the radiation field of \virgoa{} is more dilute than \ngc{}
(Fig.~\ref{fig:sed}), the ionization structure of the plasma, and line
emissivities are much more modestly affected.
The emissivity boost at 1\unit{kpc} amounts to about 2 for the \ironXXIVline{}
line, and about one order of magnitude for the $^{27}$\ion{Al}{13} line.
These are the two brightest lines in Virgo, according to
Table~\ref{table:virgo-bright-lines}.

\par
\begin{landscape}
\begin{figure*}
	\begin{centering}
		\includegraphics[scale=0.625]{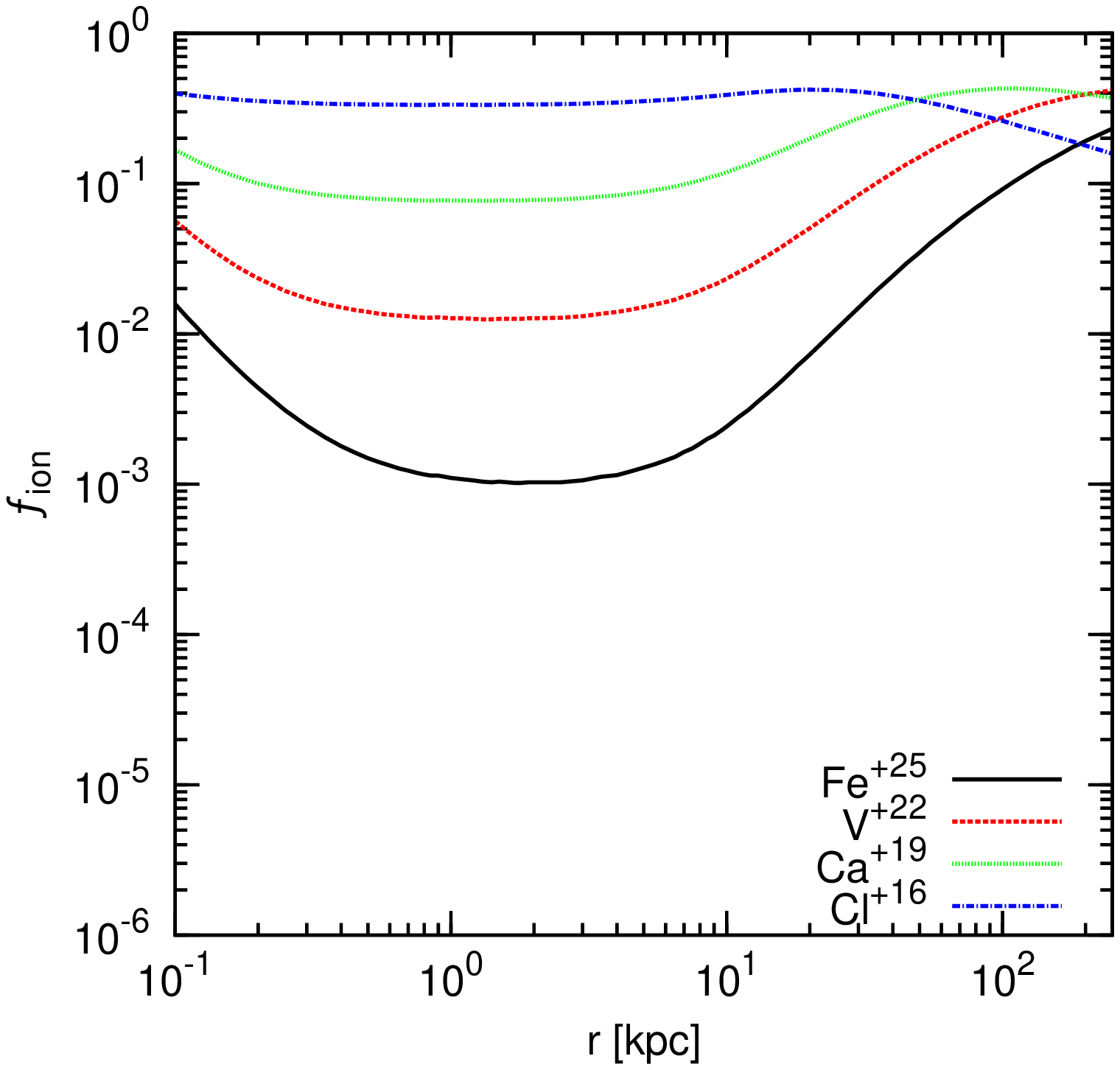}
		\includegraphics[scale=0.625]{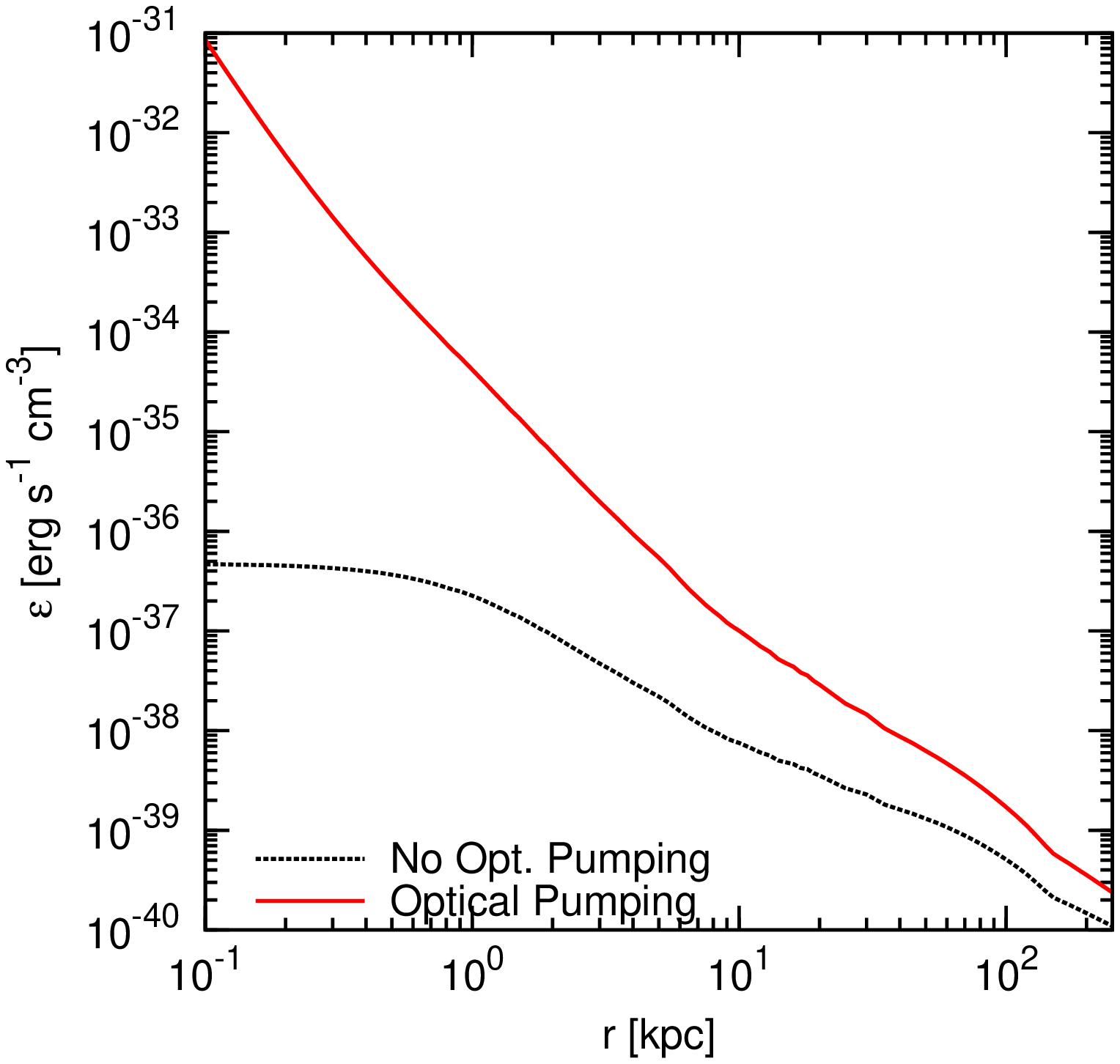}
		\caption{
			Virgo cluster results:
			{\em (Left panel)} Ionization fractions for several hydrogenic ions
			with $Z \ge 17$.  The fractions are constant within the central
			2\unit{kpc} in the absence of the Virgo A radiation.
			{\em (Right panel)} Impact of the AGN radiation field on the emissivity
			of the $^{57}$\ion{Fe}{26} $\lambda$348\unit{\mu m} line.
			Even though the AGN is not as bright as Perseus A, the emissivity
			boost is at least one order of magnitude within the central 10\unit{kpc}.
			\label{fig:virgo-ion-emis}
			}
	\end{centering}
\end{figure*}
\end{landscape}

However, the radiation field may still affect the ion fractions
of hydrogenic species with $Z \geq 17$, as shown in the left
panel of Figure~\ref{fig:virgo-ion-emis}.
In addition, the emissivity of the $^{57}$\ion{Fe}{26} line
is boosted by several orders of magnitude in the vicinity of
the AGN (right panel in Fig.~\ref{fig:virgo-ion-emis}), and
by a factor of 2 at the cluster outskirts.

\par
\begin{figure*}
	\begin{centering}
		\includegraphics[scale=0.8]{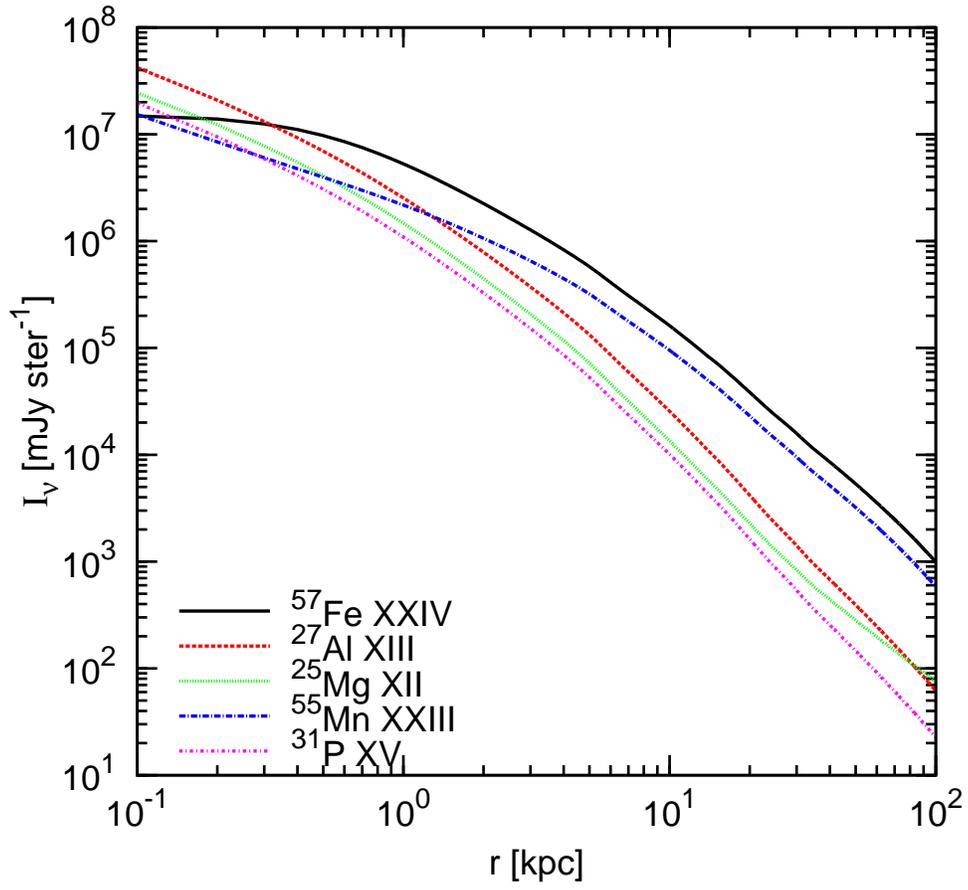}
		\caption{
			Intensity profiles for the five brightest lines in
			the Virgo cluster.
			\label{fig:virgo-profiles}
			}
	\end{centering}
\end{figure*}

The intensity profiles of the five brightest lines
are shown in Figure~\ref{fig:virgo-profiles}.
While their central values are comparable, the intensities
at 5\% of the virial radius (12\unit{kpc}) vary by almost 2
orders of magnitude.
This behavior is the opposite of that in Perseus.

\par
We have also examined the emissivity that arises from the arms of
cool gas emanating from the AGN \citep{Forman-Virgo-2007}.
We have used the emission measure maps of \citet{Werner-Virgo-2010},
along with their estimates for the line-of-sight depth of the arms
for various values of the inclination angle from the sightline.
Our calculations suggest that due to the limited depth along the
sightline, the arms increase line intensities by up to 1\%.


\begin{landscape}
\begin{deluxetable}{lcccccccc}
\tabletypesize{\rm\small}
\tablecolumns{9}
\tablewidth{0pt}
\tablecaption{Virgo Brightest Lines\label{table:virgo-bright-lines}}
\tablehead
{
    Species                     &
    $\lambda$                   &
    I$_\nu^\mathrm{max}$        &
    T$_b^\mathrm{max}$          &
    I$_\nu^\mathrm{apr}$        &
    T$_b^\mathrm{apr}$          &
    I$_\nu^\mathrm{tot}$        &
    T$_b^\mathrm{tot}$          &
    F / F$_\mathrm{cont}$    \\
                                &
    (cm)                        &
    (mJy ster$^{-1}$)           &
    (mK)                        &
    (mJy ster$^{-1}$)           &
    (mK)                        &
    (mJy ster$^{-1}$)           &
    (mK)                        &
                             \\
    (1) &
    (2) &
    (3) &
    (4) &
    (5) &
    (6) &
    (7) &
    (8) &
    (9)
}
\startdata

$^{57}$\ion{Fe}{24}  	& 3.068E$-$01	& 8.388E+06	& 2.8553E$-$02	& 4.343E+06	& 1.4782E$-$02	& 1.220E+03	& 4.1524E$-$06	& 1.352E$-$08	\\
$^{27}$\ion{Al}{13}  	& 1.210E$-$02	& 2.378E+07	&        ---	& 3.636E+06	&        ---	& 2.452E+02	&        ---	& 4.162E$-$07	\\
$^{25}$\ion{Mg}{12}  	& 6.540E$-$02	& 1.377E+07	&        ---	& 2.131E+06	&        ---	& 1.488E+02	&        ---	& 4.689E$-$08	\\
$^{55}$\ion{Mn}{23}  	& 1.520E$-$02	& 8.652E+06	&        ---	& 1.992E+06	&        ---	& 6.828E+02	&        ---	& 1.275E$-$07	\\
$^{31}$\ion{P}{15}   	& 1.510E$-$02	& 1.113E+07	&        ---	& 1.623E+06	&        ---	& 1.006E+02	&        ---	& 1.389E$-$07	\\
$^{29}$\ion{Si}{14}  	& 3.790E$-$02	& 1.078E+07	&        ---	& 1.616E+06	&        ---	& 9.750E+01	&        ---	& 5.697E$-$08	\\
$^{14}$\ion{N}{7}    	& 5.652E$-$01	& 1.271E+06	& 1.4688E$-$02	& 8.449E+05	& 9.7607E$-$03	& 1.333E+02	& 1.5403E$-$06	& 2.880E$-$09	\\
$^{23}$\ion{Na}{11}  	& 2.960E$-$02	& 4.855E+06	&        ---	& 7.471E+05	&        ---	& 5.494E+01	&        ---	& 3.787E$-$08	\\
$^{59}$\ion{Co}{25}  	& 9.150E$-$03	& 2.800E+06	&        ---	& 5.349E+05	&        ---	& 1.901E+02	&        ---	& 5.491E$-$08	\\
$^{33}$\ion{S}{16}   	& 3.270E$-$02	& 2.682E+06	&        ---	& 3.775E+05	&        ---	& 2.648E+01	&        ---	& 1.446E$-$08	\\
$^{35}$\ion{Cl}{17}  	& 2.130E$-$02	& 2.606E+06	&        ---	& 3.452E+05	&        ---	& 2.833E+01	&        ---	& 1.971E$-$08	\\
$^{53}$\ion{Cr}{22}  	& 1.130E$-$01	& 9.341E+05	& 4.3137E$-$04	& 2.131E+05	& 9.8419E$-$05	& 5.368E+01	& 2.4789E$-$08	& 1.868E$-$09	\\
$^{39}$\ion{K}{17}   	& 2.940E$-$01	& 3.290E+05	& 1.0284E$-$03	& 1.760E+05	& 5.5011E$-$04	& 4.377E+01	& 1.3683E$-$07	& 6.916E$-$10	\\
$^{39}$\ion{K}{19}   	& 3.170E$-$02	& 1.487E+06	&        ---	& 1.658E+05	&        ---	& 1.777E+01	&        ---	& 6.026E$-$09	\\
$^{37}$\ion{Cl}{17}  	& 2.550E$-$02	& 8.809E+05	&        ---	& 1.166E+05	&        ---	& 9.491E+00	&        ---	& 5.412E$-$09	\\

\enddata
\end{deluxetable}
\end{landscape}

\subsection{Extended Orion Nebula}\label{sec:orion}

\par
\citet{Orion-XMM} discovered two bubbles of hot gas in the Extended Orion Nebula,
southwest of the Huygens region.
These bubbles are thought to be the wind of the Trapezium stars, which is being
channeled into the Orion-Eridanus Superbubble, a sparse structure of hot plasma,
elongated about 100\unit{kpc} along the sightline, and seen to the south of the
Orion Nebula.
Of particular relevance to our discussion is the discovery of radioactivity
due to the rare $^{26}$Al in the Superbubble \citep{Diehl-26Al-Orion}, which
supports the scenario of replenishment over short time scales, and suggests
the presence of other isotopes at significant fractions.

\par
We have modeled the North and South cavities adopting their best-fit physical
conditions and an average path length of 2\unit{pc}.
The density and temperature for the North cavity are 0.47\unit{\pcc}, and 1.73\unit{MK},
respectively.
For the South cavity, the physical conditions are 0.23\unit{\pcc}, and 2.08\unit{MK}.

\par
The results of these calculations are presented in Table~\ref{table:EON-lines}.
Column 5 lists the line surface brightness in units of the surface brightness
of the continuum (shown in the top right panel of Fig.~\ref{fig:xray-complex}).
Both lines arise from species whose ionization fractions peak around 2\unit{MK}.

\par
Note that rough estimates for these lines may also be obtained by using the contour
plots of Figure~\ref{fig:coronal-lines-norm}.
Firstly, we calculate the continuum surface brightness.
From Figure~3 of \citet{Orion-XMM}, we approximate the 0.4--0.7\unit{keV} count
rate as 0.12 and 0.3\unit{counts \, s^{-1}} for the North and South cavities,
respectively.
We assume that the average energy of each photon in this band is 0.5\unit{keV},
and take the relevant effective area of the PN detector aboard \xmm{} to be
1000\unit{cm^2}.
We adopt the parallactic distance to Orion \citep[410\unit{pc};][]{Dist-parall}
to convert the projected cavity areas to solid angles.
The continuum surface brightness turns out to be $\sim$7E$-$8 and
$\sim$3E$-8$\unit{erg \, s^{-1} \, cm^{-2} \, ster^{-1}}, for the North and South
cavities.

\par
The two brightest among the lines of Figure~\ref{fig:coronal-lines-norm}
for the plasma conditions in the two cavities are indeed $^{27}$\ion{Al}{11}
$\lambda$1.206\unit{mm}, and $^{14}$\ion{N}{7} $\lambda$5.625\unit{mm}, with respective
surface brightness of 1E$-$6 and 1E$-$7 relative to the continuum.
These rough estimates are in factor of 3 agreement with the results listed in 
Column~5 of Table~\ref{table:EON-lines}.
The frequency broadening for these lines is $\sim$50\unit{MHz}, and $\sim$14\unit{MHz},
leading to intensities within factors of a few from the values listed in Column~3 of
Table~\ref{table:EON-lines}.
These discrepancies are in part due to the very approximate continuum surface brightness
values adopted here.
However, our approach illustrates that the surface brightness ratios of
Figure~\ref{fig:coronal-lines-norm} provide a convenient alternative to
modeling when rough estimates of optically thin line intensities are desired.

\par
We have also modeled the emission that arises from the Superbubble itself, adopting
for the density, temperature, and sightline depth the values 0.015\unit{\pcc},
3.25\unit{MK}, and 150\unit{kpc}, respectively \citep{Guo-Eridanus-Superbubble}.
However, the signal remains below 1\unit{\mu K} and is not discussed further here.
In fact, the signal would remain below that limit even if the isotope ratios were
enhanced as expected for Wolf-Rayet stellar winds \citep{Wolf-Rayet-isotopes}.

\begin{deluxetable}{lcccc}
\tabletypesize{\rm\small}
\tablecolumns{5}
\tablewidth{0pt}
\tablecaption{EON Brightest Lines \label{table:EON-lines}}
\tablehead
{
    Species               &
    $\lambda$             &
    I$_\nu$		  &
    T$_b$                 &
    I / I$_\mathrm{cont}$\\
                          &
    (cm)                  &
    (mJy ster$^{-1}$)     &
    (mK)                  &
                         \\
    (1) &
    (2) &
    (3) &
    (4) &
    (5)
}
\startdata

\multicolumn{5}{c}{{\em EON North}}	\\ \hline
$^{27}$\ion{Al}{11}        	& 1.206E$-$01	& 3.327E+05	& 1.7525E$-$04	& 1.0333E$-$06	\\
$^{14}$\ion{N}{7}          	& 5.652E$-$01	& 3.164E+05	& 3.6604E$-$03	& 2.9118E$-$07	\\

\hline\\

\multicolumn{5}{c}{{\em EON South}}	\\ \hline
$^{27}$\ion{Al}{11}        	& 1.206E$-$01	& 1.032E+05	& 5.4344E$-$05	& 5.5171E$-$07	\\
$^{14}$\ion{N}{7}          	& 5.652E$-$01	& 9.846E+04	& 1.1390E$-$03	& 1.5601E$-$07	\\

\enddata
\end{deluxetable}

\section{Discussion \& Conclusions}\label{sec:discussion}

\subsection{Galaxy Clusters}

\par
\citet{SunyaevChurazov1984} and D'Cruz \& Sarazin (Sarazin, private communication, 2013)
predicted the brightness temperature of the \ironXXIVline{} line to be around 1\unit{mK}.
Our calculations draw a more pessimistic picture.
This is in part due to the more accurate description of the X-ray emitting gas in cool
cluster cores, obtained with recent X-ray observations.
In particular, in previous work on Perseus the density has likely been overestimated
by a factor of a few, which accounts for at least one order of magnitude of the
discrepancy, in the absence of optical pumping.
Even when the effects of the AGN radiation are incorporated, the maximum intensity
remains well below previous predictions.

\par
Our calculations suggest that the \ironiso{} line will be brighter in Virgo than in Perseus,
with a brightness temperature in our fiducial aperture of only 15\unit{\mu K}.
A 3--$\sigma$ detection would require about $\sim$600~hour exposure with the full Atacama
Large Millimeter Array ({\em ALMA}) (T$_b \sim 20\unit{\mu K}$ within an 8\arcsec{} beam).


\par
Our calculations suggest that the \ironXXIVline{} line is not always
the brightest hyperfine line, as suggested by previous work.
Both clusters produce a number of lines (\ion{Al}{13} $\lambda$121\unit{\mu m},
\ion{P}{15} $\lambda$151\unit{\mu m}, and \ion{Mn}{23} $\lambda$152\unit{\mu m})
with intensities within a factor of 3 of the \ironXXIVline{} line.
In fact, in Perseus these sub-mm lines are brighter.
Clearly, our understanding of nucleosynthesis in galaxy clusters would benefit
from sub-mm spectrometers.

\par
Clusters at higher redshift are not necessarily more likely to be observed.
Because the majority of emission arises from the cluster core, the brightness
temperature of the entire cluster drops dramatically by about 4--5 orders of
magnitude when the integration extends over the entire cluster volume.
An optimal solid angle will depend on the line intensity profile, as well as
on the cluster redshift.

\par
Interestingly, we find that the AGN can strongly affect the plasma ionization
structure within the central 1\unit{kpc}.
The density of high ionization stages is enhanced, leading to a reduction
of line emission due to lower stages.
This occurs only for high-$Z$ elements about M87.
It is possible that this effect could have important implications for
the interpretation of spectra of bright AGN in the X-ray, as well as
at longer wavelengths.

\subsection{Orion Nebula}

\par
The X-ray cavities in the Extended Orion Nebula are thought
to have formed by hot plasma leaking out of the Huygens region
of the Orion Nebula.
Hyperfine line detections in these regions would help constrain
isotopic ratios in the molecular cloud, and could shed some
light into nucleosynthesis in OB stars.
However, both cavities in the EON are quite faint, with the
North cavity being about 3 times brighter than the South.
The brightest line in terms of intensity is \ion{Al}{11}
$\lambda$1.206\unit{mm}, although in terms of brightness
temperature the \ion{N}{7} line is substantially more
promising.
The brightness temperature of this line is about 3\unit{\mu K},
comparable to Perseus.
This line is inaccessible to ground-based telescopes due to
atmospheric opacity.

\section{Summary}\label{sec:summary}

\par
We have investigated the possibility of observing the hyperfine lines
of \citet{SunyaevChurazov1984} that arise from high ionization species
in hot plasmas.

\par
Our calculations are carried out with the spectral synthesis code \cloudy{},
and incorporate direct, resonant, and indirect excitations due to collisional
and radiative processes.
In addition, radiative and dielectronic recombinations are included
self-consistently.
The framework provided by \cloudy{} allows for the uniform treatment
of all hyperfine lines.

\par
Our calculations incorporate stimulated radiative transitions by the
CMB, with corrections performed according to \citet{Chatzikos-pumping}.
We find that bright point sources, such as Perseus A, can significantly
alter the ionization structure of a plasma at close proximity, as well
line emissivity at larger distances.
Less bright point sources, such as Virgo A, can affect the ionization
fraction of highly ionized heavy elements, and the emissivities of
lines with wavelengths in the range 200--1100\unit{\mu m}.
It may be the case that the AGN radiation affects optical lines, as well.
If this turns out to be true, it will likely affect the interpretation
of the nuclear spectra of Perseus~A, as well as of other systems.

\par
We have computed hyperfine line intensities over grids
covering a wide range of density and temperature, as an
aid to observations.
We have also expressed these line intensities in units of
the X-ray continuum in the range 18--28\unit{\AA}, which
includes the line complex due to carbon, nitrogen and oxygen
ions.

\par
We have improved upon the estimates of \citet{SunyaevChurazov1984}
and D'Cruz \& Sarazin \citep{Sarazin-priv-2013} on the brightness
of the \ironXXIVline{} line.
Our results are not encouraging: in Perseus the line is more
than 2 orders of magnitude fainter than previously thought.
The line is brighter in Virgo by a factor of a few, but,
having a brightness temperature of $\sim$20\unit{\mu K},
it would still require long exposure times with \alma{}
even for a marginal detection.
Our theoretical calculations are consistent with the observationally
motivated estimates of \citet{Liang1997}.

\par
In addition, sub-mm lines with intensities within a factor 3--4 of the iron
line are produced by both clusters, which may provide alternatives to
hyperfine line detections.

\par
Similar conclusions hold for the Extended Orion Nebula.
The high density of the cavity plasma is balanced by the low sightline
depth, leading to brightness temperatures comparable to those in Perseus.
However, the brightest line is the \ion{N}{7} $\lambda$5.625\unit{mm}
line which is inaccessible from ground-based observatories.

\par
Obviously, our understanding of nucleosynthesis, and
turbulent velocity fields in galaxy clusters and molecular
clouds would benefit from sub-mm instrumentation to resolve
these hyperfine emission lines, as well as from mm space
observatories to detect the aforementioned \ion{N}{7} line.
Our results, then, echo in part the conclusions of \citet{SunyaevDocenko2007},
who find $^{14}$\ion{N}{7} $\lambda$5.625\unit{mm} to be the best hyperfine
tracer of hot plasmas in the Milky Way.

\clearpage
\appendix
\section{Improved Fits to Direct And Resonant Excitation Data}\label{sec:app:fitsZS}

\par
In this section we present our improvements upon the fits of \citet[GF03]{Goddard2003}
to the ZS00 data.
Overall, with the present approach the ZS00 total collision strengths may be approximated
to 10\% or better for temperatures $10^6$~K or higher, and 25\% for temperatures
in the range $10^5$--$10^6$~K.
For comparison, GF03 report approximations of 30\% or better.

\par
While GF03 fit the total effective collision strengths as a function of temperature,
we model the contribution of each excitation mechanism (direct, and resonant through $n=2$,
and $n=3$) to the total collision strength of each species and isosequence independently.

\par
All temperature distributions may be fit with a log-normal function of the form
\begin{equation}
	Y^\mathrm{eff}(T)=	N \, \Bigl(\frac{T}{10^6 \, T_\mathrm{s1}}\Bigr)^{C_1}
				\exp\Biggl\{- \bigg| \frac{\log_{10}(T/(10^6 \, T_\mathrm{s1}))}{T_\mathrm{s2}} \bigg| ^{C_2} \Biggr\}	\,	,
	\label{eqn:lognorm}
\end{equation}
adopting a fiducial error of 3\% in the reported collision strength values.
Notice that the ZS00 data are evaluated at 11 temperature points (10 for iron),
sufficient to constrain the best-fit values of a 5-parameter functional form.
The accuracy of these fits is $\le 3$\% for the direct excitations, and
resonances through $n=2$ in Li-like isosequences,
$\le 20$\% for resonances through $n=3$ in Li-like ions,
and $\le 30$\% for resonances through $n=2$ in hydrogenic ions.
The values and errors of the best-fit parameters to each of these distributions
are presented in Table~\ref{tab:ZS00-best-fit}.

\par
The constants that appear in this expression ($N$, $T_\mathrm{s1}$, $C_1$, $T_\mathrm{s2}$,
and $C_2$) vary with the species, isosequence, and excitation mechanism under consideration.
Across an isosequence and excitation process, these constants display a strong dependence
on the nuclear charge, and occasionally a minor dependence on the nuclear spin.
This dependence on the nuclear properties has been modeled with good accuracy
across each isosequence and excitation mechanism with the use of the auxiliary
functions
\begin{eqnarray}
	f(x) & = &	A \, \Big\{ \log_{10}(x/B) \Big\}^C			\label{eqn:fit-aux-f}	\,	,	\\
	g(x) & = &	A \, \exp \Big\{ -(x/B)^C \Big\}			\label{eqn:fit-aux-g}	\,	,	\\
	h(x) & = &	A \, \exp \Big\{  (x/B)^C \Big\}			\label{eqn:fit-aux-h}	\,	,	\\
	k(x) & = &	A \, (x/B)^C \, \Big\{ \exp  (x/B) \Big\}^D		\label{eqn:fit-aux-k}	\,	,	\\
	l(x) & = &	A \, (x/B)^C \, \exp \Big\{  (x/B)^D \Big\}		\label{eqn:fit-aux-l}	\,	,	\\
	m(x) & = &	A \, (x/B)^C \, \exp \Big\{ -(x/B)^D \Big\}		\label{eqn:fit-aux-m}	\,	,
\end{eqnarray}
as shown in Table~\ref{tab:ZS00-best-fit-corr}.
Notice that these forms are not independent of each other,
but are employed here for convenience.
While they are constrained by the five Li-like species,
the two hydrogenic species are not sufficient to constrain
the fits in any reliable fashion.

\par
In the Li-like ions, parameters with undefined uncertainties have been held fixed to improve
the goodness-of-fit measure, as it was found that their exact values did not significantly
affect the agreement of the best-fit function with the available data points.

\par
By contrast, to circumvent the lack of sufficient data in the hydrogenic sequences,
it was assumed that the scaling ought to be similar to that of the Li-like isosequence.
In detail, we have assumed that the scaling with nuclear properties (parameter $B$) is
the same across isosequences, allowing only the normalization and power-law exponent
(parameters $A$ and $C$) to vary.
When present, the power-law associated with the exponential (parameter $D$)
was held fixed at the Li-like value.

\par
Notice also that owing to the similarity in the shape and amplitude of their profiles,
we have adopted the functional forms of $n=3$ resonances of Li-like ions for the $n=2$
resonances of the hydrogenic ions.

\par
\begin{figure*}
	\begin{centering}
		\includegraphics[scale=0.7]{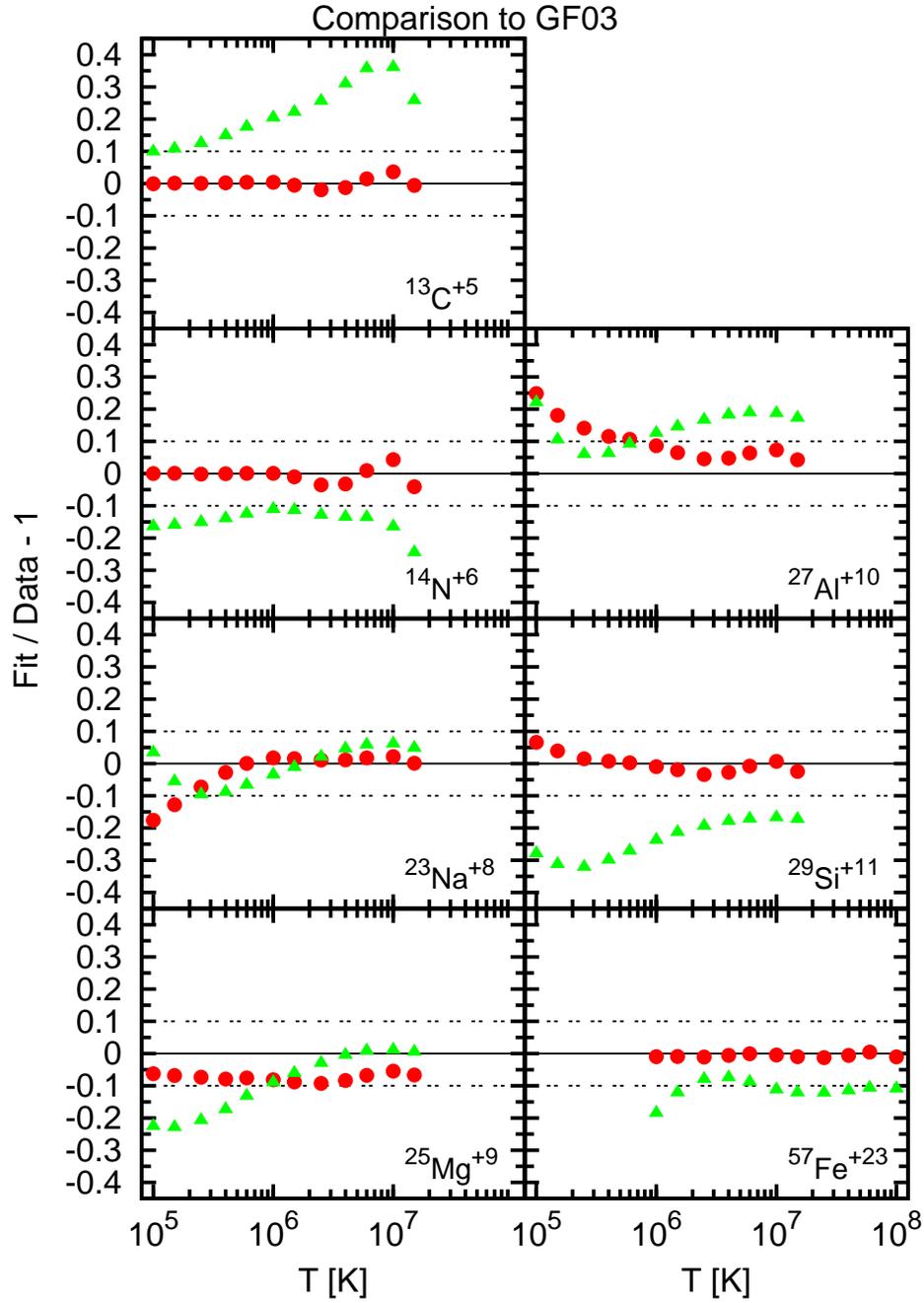}
		\caption{
			Comparison of the fit residuals to the total effective
			collision strength of the ZS00 data for the present work
			(red solid circles) and the work of GF03 (green solid triangles).
			The dashed lines indicate $\pm 10$\% deviations
			from the data.
			\label{fig:comp-GF03}
			}
	\end{centering}
\end{figure*}

A comparison between the present work and the fits of GF03 is presented
in Figure~\ref{fig:comp-GF03}.
In all but two cases, the fitting function employed in the present
work results in at most a 10\% deviation from the exact data.
The exceptions are the Li-like sodium and aluminum ions, which possess
20--30\% deviations at the low end of the temperature range.
Overall, the present approach results in more accurate fits than GF03,
with the exception of the Li-like magnesium ion at temperatures greater
than 10$^6$~K.

\par
These fits have been implemented in \cloudy{}
as an update to the existing data of GF03.
Note that in computing the collision strengths for the \citet{SunyaevChurazov1984}
hydrogenic, and lithium-like ions, we have imposed an upper limit ($C_1 \leq 1$)
on the power-law value for hydrogenic resonances, and limits on the power-law and
temperature scaling values ($C_1 \geq -1$ and $T_{s1} \leq 35$) for Li-like $n=3$
resonances, respectively, to prevent divergences to unphysical values.
We adopt the lithium-like profiles for the sodium-like ions for lack of
any numerical data.
Such inadequacies illustrate the need for additional calculations
primarily for hydrogenic and sodium-like ions.

\begin{landscape}
\begin{deluxetable}{lcccccccccccc}
\tabletypesize{\rm\scriptsize}
\tablecolumns{13}
\tablewidth{0pt}
\tablecaption{Best-Fit Parameters to ZS00 Data\label{tab:ZS00-best-fit}}
\tablehead
  {
    Ion				&
    $N_\mathrm{dof}$		&
    $\chi_\mathrm{dof}^2$	&
    $N$				&
    $\sigma_N$			&
    $T_\mathrm{s1}$		&
    $\sigma_{T_\mathrm{s1}}$	&
    $C_1$			&
    $\sigma_{C_1}$		&
    $T_\mathrm{s2}$		&
    $\sigma_{T_\mathrm{s2}}$    &
    $C_2$			&
    $\sigma_{C_2}$		\\ 
 (1) &
 (2) &
 (3) &
 (4) &
 (5) &
 (6) &
 (7) &
 (8) &
 (9) &
 (10) &
 (11) &
 (12) &
 (13) 
  }
\startdata

%
%
\hline
\multicolumn{13}{c}{{Direct (Background) Excitations}} \\
\hline
$^{13}$C$^{+5}$		& 7 &	0.00313728 &	0.0463139	& 0.0002839	& 0.207503	& 0.02996	& -0.0475049	& 0.00721	& 1.86432	& 0.04121	& 2.56332	& 0.04545	\\
$^{14}$N$^{+6}$		& 7 &	0.00155694 &	0.0610659	& 0.0002067	& 0.223362	& 0.02193	& -0.039488	& 0.003979	& 1.93797	& 0.03046	& 2.67975	& 0.03559	\\	
$^{23}$Na$^{+8}$	& 7 &	0.0126283  &    0.0735341 	& 0.002088	& 0.178673	& 0.06534	& -0.0816228	& 0.03259	& 1.68484	& 0.06735	& 2.27324	& 0.06218	\\	
$^{25}$Mg$^{+9}$	& 7 &	0.00893359 &	0.0928638 	& 0.001721	& 0.185271	& 0.05118	& -0.0715463	& 0.02126	& 1.72992	& 0.05934	& 2.34265	& 0.05738	\\	
$^{27}$Al$^{+10}$	& 7 &	0.00585586 &	0.0768643	& 0.0009545	& 0.191162	& 0.04052	& -0.0633176	& 0.01421	& 1.77343	& 0.05083	& 2.40811	& 0.05123	\\	
$^{29}$Si$^{+11}$	& 7 &	0.00314582 &	0.0166136 	& 0.0001278	& 0.198736	& 0.02943	& -0.056951	& 0.008833	& 1.81119	& 0.03837	& 2.4691	& 0.04003	\\	
$^{57}$Fe$^{+23}$	& 6 &	0.0139415  &    0.00374383	& 0.0001595	& 1.78522	& 0.7198	& -0.109575	& 0.04304	& 1.61725	& 0.0603	& 2.21178	& 0.05548	\\	
%
%
\hline
\multicolumn{13}{c}{{$n=2$ Resonant Excitations}} \\
\hline
$^{13}$C$^{+5}$		& 7 & 	60.4757 	& 0.00286807	& 0.0003172	& 3.42741	& 0.4091	& 0.336739	& 0.1964	& 0.627927	& 0.03002	& 3.65514	& 0.1326 	\\       
$^{14}$N$^{+6}$		& 7 & 	106.616		& 0.00481214	& 0.0007836	& 3.36366	& 0.3964 	& 0.801919	& 0.2618 	& 0.577669	& 0.02999 	& 3.65531 	& 0.1519 	\\ 
$^{23}$Na$^{+8}$	& 7 & 	0.0233641	& 0.126021	& 0.01115	& 3.79216	& 0.3447	& -0.96184 	& 0.004376 	& 1.64596	& 0.03367	& 3.56222	& 0.06541	\\	
$^{25}$Mg$^{+9}$	& 7 & 	0.00987751	& 0.157279	& 0.01532	& 4.34866 	& 0.4322 	& -0.972123	& 0.003537 	& 1.81967	& 0.03689	& 3.32988	& 0.06   	\\	
$^{27}$Al$^{+10}$	& 7 & 	0.0329427	& 0.133248	& 0.01365	& 3.95003	& 0.4163	& -0.959629	& 0.005621 	& 1.61544	& 0.03793	& 3.45352	& 0.07018	\\	
$^{29}$Si$^{+11}$	& 7 & 	0.0201666  	& 0.0344953	& 0.002928	& 3.95752	& 0.3462	& -0.958917	& 0.004656 	& 1.65588	& 0.03101 	& 3.34559	& 0.05451	\\	
$^{57}$Fe$^{+23}$	& 6 & 	0.00708709	& 0.0146527	& 0.0007875	& 1.88574	& 0.1189	& -0.836432	& 0.008784 	& 2.67792	& 0.08321	& 1.44764	& 0.02205	\\
%
%
\hline
\multicolumn{13}{c}{{$n=3$ Resonant Excitations}} \\
\hline
$^{23}$Na$^{+8}$ 	& 7 &	4.9649 	   &	0.0216931 	& 0.00228 	& 3.74735 	& 0.5855 	& -0.561302	& 0.07659	& 0.86893	& 0.03629	& 3.19419	& 0.09567	\\	
$^{25}$Mg$^{+9}$ 	& 7 &	8.8006	   &	0.0330297 	& 0.003101 	& 3.47036 	& 0.5493	& -0.442042	& 0.09474	& 0.831089	& 0.03501	& 3.24333	& 0.1095 	\\       
$^{27}$Al$^{+10}$	& 7 &	13.1543	   &	0.0313274 	& 0.002508 	& 3.23321	& 0.4854	& -0.31384	& 0.1071	& 0.800671	& 0.03228	& 3.31426	& 0.1206 	\\       
$^{29}$Si$^{+11}$	& 7 &	17.0828	   &	0.00739256 	& 0.0005303	& 3.0921	& 0.4331 	& -0.193367	& 0.1142	& 0.778544	& 0.03019	& 3.39744	& 0.1274 	\\       
$^{57}$Fe$^{+23}$	& 6 &	0.925002   &	0.00115868	& 0.0001332	& 31.1566	& 4.592 	& -0.731262	& 0.04629	& 0.965965	& 0.03518	& 2.95329	& 0.07596	

\enddata

\end{deluxetable}
\end{landscape}

\begin{landscape}
\begin{deluxetable}{lccccccccccc}
\tabletypesize{\rm\scriptsize}
\tablecolumns{12}
\tablewidth{0pt}
\tablecaption{Correlations of Best-Fit Parameters to ZS00 Data\label{tab:ZS00-best-fit-corr}}
\tablehead
  {
    Parameter    		&
    Function     		&
    $N_\mathrm{dof}$    	&
    $\chi_\mathrm{dof}^2$	&
    A    			&
    $\sigma_A$    		&
    B    			&
    $\sigma_B$    		&
    C    			&
    $\sigma_C$    		&
    D    			&
    $\sigma_D$   \\
 (1) &
 (2) &
 (3) &
 (4) &
 (5) &
 (6) &
 (7) &
 (8) &
 (9) &
 (10) &
 (11) &
 (12)   
  }
\startdata

\hline
\multicolumn{12}{c}{{Li-like Direct (Background) Excitations}} \\ \hline
 N    &
 $f(\sqrt{I}/Z)$    &
 2    &
 7.308e+00    &
 1.286e-02    &
 1.850e-02    &
 4.481e-03    &
 2.954e-03    &
 5.03933    &
 1.215    &
 \ldots    &
 \ldots \\

 $T_{s1}$    &
 $h(Z)$    &
 2    &
 4.266e-04    &
 1.708e-01    &
 1.821e-03    &
 2.143e+01    &
 1.383e-01    &
 4.41718    &
 0.1231    &
 \ldots    &
 \ldots \\

 $C_1$    &
 $l(Z)$    &
 1    &
 3.519e-05    &
 -2.339e-02    &
 5.039e-04    &
 2.404e+01    &
 3.609e-01    &
 -1.59178    &
 0.05401    &
 6.54585    &
 0.8688 \\

 $T_{s2}$    &
 $m(Z)$    &
 2    &
 2.650e-04    &
 2.364e+00    &
 7.109e-02    &
 3.125e+01    &
 9.468e-01    &
 0.323085    &
 0.01819    &
 6.1953    &
 1.305 \\

 $C_2$    &
 $m(Z)$    &
 1    &
 5.809e-06    &
 3.217e+00    &
 9.603e-03    &
 2.969e+01    &
 1.390e-01    &
 0.349369    &
 0.001352    &
 8.3886    &
 0.3214 \\\hline

\multicolumn{12}{c}{{Li-like $n=2$ Resonances}} \\ \hline
 N    &
 $f(I/Z)$    &
 2    &
 8.700e-01    &
 5.249e-02    &
 3.626e-02    &
 5.186e-03    &
 3.286e-03    &
 2.26828    &
 0.6926    &
 \ldots    &
 \ldots \\

 $T_{s1}$    &
 $g(Z)$    &
 2    &
 6.460e-01    &
 4.743e+00    &
 3.093e-01    &
 2.678e+01    &
 \ldots    &
 2.33347    &
 0.6295    &
 \ldots    &
 \ldots \\

 $C_1$    &
 $g(Z)$    &
 2    &
 2.756e+00    &
 -9.720e-01    &
 1.524e-02    &
 4.127e+01    &
 1.192e+01    &
 4.10155    &
 2.742    &
 \ldots    &
 \ldots \\

 $T_{s2}$    &
 $\frac{Z}{\sqrt{I}} h(\sqrt{I}/Z)$    &
 2    &
 4.368e+00    &
 6.338e-02    &
 7.959e-03    &
 1.100e-01    &
 1.049e-02    &
 1.57284    &
 0.3833    &
 \ldots    &
 \ldots \\

 $C_2$    &
 $g(Z)$    &
 2    &
 2.926e+00    &
 3.569e+00    &
 2.206e-01    &
 2.664e+01    &
 2.603e-01    &
 4.19725    &
 1.812    &
 \ldots    &
 \ldots \\\hline

\multicolumn{12}{c}{{Li-like $n=3$ Resonances}} \\ \hline
 N    &
 $f(I/Z)$    &
 2    &
 1.319e+00    &
 2.717e-02    &
 1.615e-03    &
 1.597e-02    &
 1.042e-03    &
 1.25218    &
 0.1487    &
 \ldots    &
 \ldots \\

 $T_{s1}$    &
 $k(Z)$    &
 2    &
 3.740e-02    &
 7.168e-03    &
 1.646e-03    &
 1.000e+01    &
 \ldots    &
 -8.85041    &
 0.4891    &
 6.47427    &
 0.2666 \\

 $C_1$    &
 $l(Z)$    &
 2    &
 2.240e+01    &
 -3.762e-01    &
 7.083e-02    &
 1.000e+01    &
 \ldots    &
 -7.94404    &
 0.9693    &
 2.20888    &
 0.09759 \\

 $T_{s2}$    &
 $k(Z)$    &
 2    &
 4.878e-03    &
 2.933e-01    &
 6.833e-03    &
 1.000e+01    &
 \ldots    &
 -1.88971    &
 0.04903    &
 1.15294    &
 0.02704 \\

 $C_2$    &
 $k(Z)$    &
 2    &
 5.896e-02    &
 5.920e+00    &
 4.141e-01    &
 1.000e+01    &
 \ldots    &
 1.05838    &
 0.1436    &
 -0.656385    &
 0.07999 \\\hline

\multicolumn{12}{c}{{H-like Direct (Background) Excitations}} \\ \hline
 N    &
 $f(\sqrt{I}/Z)$    &
 0    &
 \ldots       &
 8.499e-03    &
 \ldots       &
 4.481e-03    &
 \ldots    &
 4.83542    &
 \ldots    &
 \ldots    &
 \ldots \\

 $T_{s1}$    &
 $h(Z)$    &
 0    &
 \ldots       &
 1.4966e-01   &
 \ldots       &
 2.143e+01    &
 \ldots       &
 0.837023     &
 \ldots       &
 \ldots       &
 \ldots	\\

 $C_1$    &
 $l(Z)$    &
 0    &
 \ldots       &
 -8.976e-03    &
 \ldots       &
 2.404e+01    &
 \ldots    &
 -1.20034    &
 \ldots    &
 6.54585    &
 \ldots \\

 $T_{s2}$    &
 $m(Z)$    &
 0    &
 \ldots       &
 2.824e+00    &
 \ldots       &
 3.125e+01    &
 \ldots    &
 0.251719    &
 \ldots    &
 6.1953    &
 \ldots \\

 $C_2$    &
 $m(Z)$    &
 0    &
 \ldots       &
 4.064e+00    &
 \ldots    &
 2.969e+01    &
 \ldots    &
 0.288187    &
 \ldots    &
 8.3886    &
 \ldots    \\ \hline

\multicolumn{12}{c}{{H-like $n=2$ Resonances}} \\ \hline
 N    &
 $f(I/Z)$    &
 0    &
 \ldots    &
 5.270e-03    &
 \ldots    &
 1.597e-02    &
 \ldots    &
 1.83289    &
 \ldots    &
 \ldots    &
 \ldots \\

 $T_{s1}$    &
 $k(Z)$    &
 0    	   &
 \ldots    &
 7.747e-03 &
 \ldots    &
 1.000e+01 &
 \ldots    &
 -4.32176  &
 \ldots    &
 6.47427   &
 \ldots \\

 $C_1$    &
 $l(Z)$    &
 0    &
 \ldots    &
 2.797e+00    &
 \ldots    &
 1.000e+01    &
 \ldots    &
 4.77743    &
 \ldots    &
 2.20888    &
 \ldots     \\

 $T_{s2}$    &
 $k(Z)$    &
 0    &
 \ldots    &
 1.627e-01    &
 \ldots    &
 1.000e+01    &
 \ldots    &
 -1.28911    &
 \ldots    &
 1.15294    &
 \ldots    \\

 $C_2$    &
 $k(Z)$    &
 0    &
 \ldots    &
 6.737e+00    &
 \ldots    &
 1.000e+01    &
 \ldots    &
 0.426109    &
 \ldots    &
 -0.656385    &
 \ldots
\enddata

\end{deluxetable}
\end{landscape}

\section{SED Compilation}\label{app:sec:seds}

\par
\begin{figure*}
	\begin{centering}
		\includegraphics[scale=0.7]{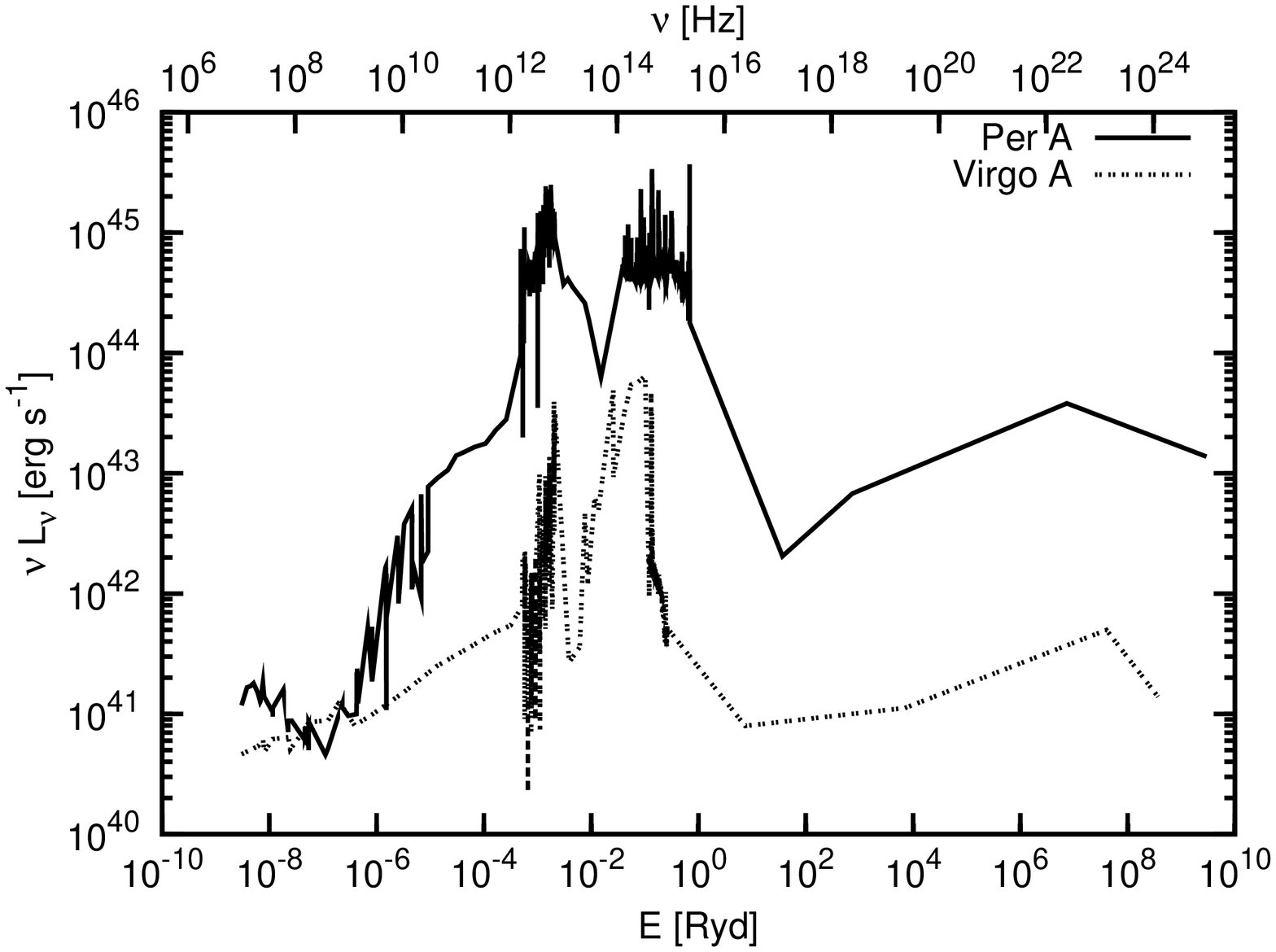}
		\caption{
			The spectral energy distribution of \ngc{} and \virgoa.
			See text for details.
			\label{fig:sed}
			}
	\end{centering}
\end{figure*}

Because the hyperfine lines considered in this paper cover a wide spectral
range (the shortest wavelength is 10\unit{\mu m}), and because indirect
radiative excitations can be important, the spectral energy distribution
(SED) over the entire electromagnetic spectrum of each source is required.
In this Appendix, we discuss how we compiled the SEDs of \ngc{} and
\virgoa{}, shown in Figure~\ref{fig:sed}.
In both cases, we have used disjoint multiband observations mainly derived
from the NASA/IPAC Extragalactic Database (NED\footnote{http://ned.ipac.caltech.edu/}).

\par
In the case of \ngc{}, the spectral segments listed in Table~\ref{tab:sed-det}
cover various epochs over the last 50 years, and are subject to the long term
variability observed for this galaxy \citep{Nesterov1995}.
This variability amounts to roughly a factor of 5 in the radio, and a factor of 2
in the optical, and it is thought to arise from accretion onto the central black
hole, as well as jet interaction with the accreting gas.
Currently, the AGN is brightening, as seen for example in the mm
observations of \citet{Trippe2011}.

\par
Obviously, arriving at an ``average'' SED requires accounting
for temporal variability.
The corrections are listed in Column 6 of Table~\ref{tab:sed-det}.
Because most observations were taken close to the lightcurve
minimum, the temporal corrections are set to the square root
of maximum variability.
When long-term monitoring is absent, the optical factor
is employed, that is, the correction is set to $\sqrt{2}$.
The values for the millimeter and submillimeter variability
are drawn from the 90\unit{GHz} and 270\unit{GHz} lightcurves
of \citet{Nesterov1995}.

\par
Aperture corrections, shown in Column 7 of Table~\ref{tab:sed-det},
are employed only for the near infrared, optical, and ultraviolet
parts of the spectrum.
The first two are obtained with long slit spectroscopy,
while the latter with grism spectroscopy.
The infrared and ultraviolet segments are adjusted
to be roughly continuous with the optical.
The joint spectrum is then corrected for reddening
following the prescription of \citet{Fitzpatrick1999}
and adopting a $R_V = 3.1$ extinction law, while its
normalization is adjusted to yield a B-filter
\citep{Bessel1990} magnitude in agreement with
\citet{Moustakas2006}.
No aperture corrections are employed with the
other segments, either because the aperture
contains the entire galaxy, or because it was
not possible to estimate the total flux (e.g., MIR).


\begin{deluxetable}{lccccccl}
\tabletypesize{\rm\small}
\tablecolumns{8}
\tablewidth{0pt}
\tablecaption{SED Compilation Details\label{tab:sed-det}}
\tablehead
  {
    Band		&
    Spectral Range	&
    Obs.\ Epoch		&
    Facility		&
    Aperture		&
    V$_\mathrm{corr}$\tablenotemark{a}	&
    A$_\mathrm{corr}$\tablenotemark{b}	&
    Ref.\tablenotemark{c}		\\
 (1) &
 (2) &
 (3) &
 (4) &
 (5) &
 (6) &
 (7) &
 (8)
  }
\startdata

Radio-CM	&
$< 30$ GHz	&
1969-2007	&
\ldots		&
\ldots		&
\ldots		&
\ldots		&
\ldots		\\
Radio-MM	&
30--857 GHz	&
2009		&
\planck{}		&
4\arcmin--33\arcmin	&
$\sqrt{5}$	&
\ldots		&
1		\\
FIR		&
43-190 \micron	&
1998		&
\iso{}		&
72\arcsec	&
$\sqrt{15}$	&
\ldots		&
2		\\
MIR		&
12--100 \micron	&
2006		&
Bok		&
75\arcsec	&
$\sqrt{2}$	&
\ldots		&
3		\\
MIR		&
6--30 \micron	&
2005		&
\spitzer{}	&
11\arcsec$\times$22\arcsec	&
$\sqrt{2}$	&
\ldots		&
4		\\
NIR		&
0.8--2.4 \micron	&
2003		&
IRTF		&
0.8\arcsec$\times$1.5\arcsec	&
$\sqrt{2}$	&
20		&
5		\\
Optical		&
0.37--0.69 \micron	&
2006		&
Bok		&
2\arcsec$\times$516\arcsec	&
$\sqrt{2}$	&
20		&
6		\\
UV		&
0.13--0.28 \micron	&
2005		&
GALEX		&
10\arcsec	&
$\sqrt{2}$	&
10		&
7		\\
X-ray soft	&
0.3--100 keV	&
2005--2007	&
\xmm{}		&
25\arcsec	&
$\sqrt{2}$	&
\ldots		&
8		\\
X-ray hard	&
0.3--100 keV	&
2005--2007	&
\swift{}		&	
17\arcmin	&
$\sqrt{2}$	&
\ldots		&
8		\\
$\gamma$-ray	&
0.2--10 GeV	&
2008		&
\fermi{}		&
2\arcdeg--0.1\arcdeg	&
$\sqrt{2}$	&
\ldots		&
9

\enddata

\tablenotetext{a}{Variability correction factor.}
\tablenotetext{b}{Aperture correction factor.}
\tablenotetext{c}{Reference List: 1: \citet{LeonTavares2012}; 2: \citet{Brauher2008};
		3: \citet{Moustakas2006}; 4: \citet{Weedman2005}; 5: \citet{Riffel2006};
		6: \citet{Buttiglione2009}; 7: GALEX coadded spectral products (http://galex.stsci.edu/GR6/);
		8: \citet{Ajello2010}; 9: \citet{Abdo2009}.}

\end{deluxetable}

\par
By contrast, the SED of \virgoa{} refers to the nuclear galaxy
region (roughly the central 35\arcsec, or 3\unit{kpc}), due to
the cluster's proximity to the Milky Way ($z = 0.00436$).
In addition, no long term variability has been observed for this
galaxy, and therefore no corrections are required.




\clearpage

\bibliography{hfs-lines,bibliography2}

\begin{thebibliography}{52}
\expandafter\ifx\csname natexlab\endcsname\relax\def\natexlab#1{#1}\fi

\bibitem[{{Abdo} {et~al.}(2009){Abdo}, {Ackermann}, {Ajello}, {Asano},
  {Baldini}, {Ballet}, {Barbiellini}, {Bastieri}, {Baughman}, {Bechtol},
  {Bellazzini}, {Blandford}, {Bloom}, {Bonamente}, {Borgland}, {Bregeon},
  {Brez}, {Brigida}, {Bruel}, {Burnett}, {Caliandro}, {Cameron}, {Caraveo},
  {Casandjian}, {Cavazzuti}, {Cecchi}, {Celotti}, {Chekhtman}, {Cheung},
  {Chiang}, {Ciprini}, {Claus}, {Cohen-Tanugi}, {Colafrancesco}, {Cominsky},
  {Conrad}, {Costamante}, {Dermer}, {de Angelis}, {de Palma}, {Digel},
  {Donato}, {do Couto e Silva}, {Drell}, {Dubois}, {Dumora}, {Farnier},
  {Favuzzi}, {Finke}, {Focke}, {Frailis}, {Fukazawa}, {Funk}, {Fusco},
  {Gargano}, {Georganopoulos}, {Germani}, {Giebels}, {Giglietto}, {Giordano},
  {Glanzman}, {Grenier}, {Grondin}, {Grove}, {Guillemot}, {Guiriec},
  {Hanabata}, {Harding}, {Hartman}, {Hayashida}, {Hays}, {Hughes},
  {J{\'o}hannesson}, {Johnson}, {Johnson}, {Johnson}, {Kadler}, {Kamae},
  {Kanai}, {Katagiri}, {Kataoka}, {Kawai}, {Kerr}, {Kn{\"o}dlseder}, {Kuehn},
  {Kuss}, {Latronico}, {Lemoine-Goumard}, {Longo}, {Loparco}, {Lott},
  {Lovellette}, {Lubrano}, {Madejski}, {Makeev}, {Mazziotta}, {McEnery},
  {Meurer}, {Michelson}, {Mitthumsiri}, {Mizuno}, {Moiseev}, {Monte},
  {Monzani}, {Morselli}, {Moskalenko}, {Murgia}, {Nakamori}, {Nolan}, {Norris},
  {Nuss}, {Ohsugi}, {Omodei}, {Orlando}, {Ormes}, {Paneque}, {Panetta},
  {Parent}, {Pepe}, {Pesce-Rollins}, {Piron}, {Porter}, {Rain{\`o}}, {Razzano},
  {Reimer}, {Reimer}, {Reposeur}, {Ritz}, {Rodriguez}, {Romani}, {Ryde},
  {Sadrozinski}, {Sambruna}, {Sanchez}, {Sander}, {Sato}, {Parkinson},
  {Sgr{\`o}}, {Smith}, {Smith}, {Spandre}, {Spinelli}, {Starck}, {Strickman},
  {Strong}, {Suson}, {Tajima}, {Takahashi}, {Takahashi}, {Tanaka}, {Taylor},
  {Thayer}, {Thompson}, {Torres}, {Tosti}, {Uchiyama}, {Usher}, {Vilchez},
  {Vitale}, {Waite}, {Wood}, {Ylinen}, {Ziegler}, {Aller}, {Aller},
  {Kellermann}, {Kovalev}, {Kovalev}, {Lister}, \& {Pushkarev}}]{Abdo2009}
{Abdo}, A.~A., {Ackermann}, M., {Ajello}, M., {et~al.} 2009, \apj, 699, 31

\bibitem[{{Ajello} {et~al.}(2010){Ajello}, {Rebusco}, {Cappelluti}, {Reimer},
  {B{\"o}hringer}, {La Parola}, \& {Cusumano}}]{Ajello2010}
{Ajello}, M., {Rebusco}, P., {Cappelluti}, N., {et~al.} 2010, \apj, 725, 1688

\bibitem[{{Asplund} {et~al.}(2009){Asplund}, {Grevesse}, {Sauval}, \&
  {Scott}}]{Asplund2009}
{Asplund}, M., {Grevesse}, N., {Sauval}, A.~J., \& {Scott}, P. 2009, \araa, 47,
  481

\bibitem[{{Bania} {et~al.}(2007){Bania}, {Balser}, {Rood}, {Wilson}, \&
  {LaRocque}}]{Bania2007}
{Bania}, T.~M., {Balser}, D.~S., {Rood}, R.~T., {Wilson}, T.~L., \& {LaRocque},
  J.~M. 2007, \apj, 664, 915

\bibitem[{{Bania} {et~al.}(1997){Bania}, {Balser}, {Rood}, {Wilson}, \&
  {Wilson}}]{Bania1997}
{Bania}, T.~M., {Balser}, D.~S., {Rood}, R.~T., {Wilson}, T.~L., \& {Wilson},
  T.~J. 1997, \apjs, 113, 353

\bibitem[{{Bessell}(1990)}]{Bessel1990}
{Bessell}, M.~S. 1990, \pasp, 102, 1181

\bibitem[{{Brauher} {et~al.}(2008){Brauher}, {Dale}, \& {Helou}}]{Brauher2008}
{Brauher}, J.~R., {Dale}, D.~A., \& {Helou}, G. 2008, \apjs, 178, 280

\bibitem[{{Buttiglione} {et~al.}(2009){Buttiglione}, {Capetti}, {Celotti},
  {Axon}, {Chiaberge}, {Macchetto}, \& {Sparks}}]{Buttiglione2009}
{Buttiglione}, S., {Capetti}, A., {Celotti}, A., {et~al.} 2009, \aap, 495, 1033

\bibitem[{{Chatzikos} {et~al.}(2013){Chatzikos}, {Ferland}, \&
  {Williams}}]{Chatzikos-pumping}
{Chatzikos}, M., {Ferland}, G.~J., \& {Williams}, R.~J.~R. 2013, ApJ, 779, 122

\bibitem[{{Churazov} {et~al.}(2004){Churazov}, {Forman}, {Jones}, {Sunyaev}, \&
  {B{\"o}hringer}}]{Churazov-XMM-Perseus}
{Churazov}, E., {Forman}, W., {Jones}, C., {Sunyaev}, R., \& {B{\"o}hringer},
  H. 2004, \mnras, 347, 29

\bibitem[{{Churazov} {et~al.}(2008){Churazov}, {Forman}, {Vikhlinin},
  {Tremaine}, {Gerhard}, \& {Jones}}]{Churazov-Virgo-2008}
{Churazov}, E., {Forman}, W., {Vikhlinin}, A., {et~al.} 2008, \mnras, 388, 1062

\bibitem[{{D'Cruz} {et~al.}(1998){D'Cruz}, {Sarazin}, \&
  {Dubau}}]{DCruzSarazin1998}
{D'Cruz}, N.~L., {Sarazin}, C.~L., \& {Dubau}, J. 1998, \apj, 501, 414 (DSD98)

\bibitem[{{Dere} {et~al.}(1997){Dere}, {Landi}, {Mason}, {Monsignori Fossi}, \&
  {Young}}]{Dere.K97CHIANTI---an-atomic-database-for-emission}
{Dere}, K.~P., {Landi}, E., {Mason}, H.~E., {Monsignori Fossi}, B.~C., \&
  {Young}, P.~R. 1997, \aaps, 125, 149

\bibitem[{{Dickey} \& {Lockman}(1990)}]{DickeyLockman1990}
{Dickey}, J.~M., \& {Lockman}, F.~J. 1990, \araa, 28, 215

\bibitem[{{Diehl}(2002)}]{Diehl-26Al-Orion}
{Diehl}, R. 2002, \nar, 46, 547

\bibitem[{{Ettori} {et~al.}(1998){Ettori}, {Fabian}, \&
  {White}}]{Ettori-ROSAT-Perseus}
{Ettori}, S., {Fabian}, A.~C., \& {White}, D.~A. 1998, \mnras, 300, 837

\bibitem[{{Fabian} {et~al.}(2006){Fabian}, {Sanders}, {Taylor}, {Allen},
  {Crawford}, {Johnstone}, \& {Iwasawa}}]{Fabian-Perseus-2006}
{Fabian}, A.~C., {Sanders}, J.~S., {Taylor}, G.~B., {et~al.} 2006, \mnras, 366,
  417

\bibitem[{{Ferland} {et~al.}(2013){Ferland}, {Porter}, {van Hoof}, {Williams},
  {Abel}, {Lykins}, {Shaw}, {Henney}, \& {Stancil}}]{CloudyReview13}
{Ferland}, G.~J., {Porter}, R.~L., {van Hoof}, P.~A.~M., {et~al.} 2013, Revista
  Mexicana de Astronomia y Astrofisica, 49, 137

\bibitem[{{Fitzpatrick}(1999)}]{Fitzpatrick1999}
{Fitzpatrick}, E.~L. 1999, \pasp, 111, 63

\bibitem[{{Forman} {et~al.}(2007){Forman}, {Jones}, {Churazov}, {Markevitch},
  {Nulsen}, {Vikhlinin}, {Begelman}, {B{\"o}hringer}, {Eilek}, {Heinz},
  {Kraft}, {Owen}, \& {Pahre}}]{Forman-Virgo-2007}
{Forman}, W., {Jones}, C., {Churazov}, E., {et~al.} 2007, \apj, 665, 1057

\bibitem[{{Garstang}(1995)}]{Garstang1995}
{Garstang}, R.~H. 1995, \apj, 447, 962

\bibitem[{{Goddard} \& {Ferland}(2003)}]{Goddard2003}
{Goddard}, W.~E., \& {Ferland}, G.~J. 2003, \pasp, 115, 647

\bibitem[{{Gould}(1994)}]{Gould1994}
{Gould}, R.~J. 1994, \apj, 423, 522

\bibitem[{{Grevesse} {et~al.}(2010){Grevesse}, {Asplund}, {Sauval}, \&
  {Scott}}]{Grevesse2010}
{Grevesse}, N., {Asplund}, M., {Sauval}, A.~J., \& {Scott}, P. 2010, \apss,
  328, 179

\bibitem[{{G{\"u}del} {et~al.}(2008){G{\"u}del}, {Briggs}, {Montmerle},
  {Audard}, {Rebull}, \& {Skinner}}]{Orion-XMM}
{G{\"u}del}, M., {Briggs}, K.~R., {Montmerle}, T., {et~al.} 2008, Science, 319,
  309

\bibitem[{{Guo} {et~al.}(1995){Guo}, {Burrows}, {Sanders}, {Snowden}, \&
  {Penprase}}]{Guo-Eridanus-Superbubble}
{Guo}, Z., {Burrows}, D.~N., {Sanders}, W.~T., {Snowden}, S.~L., \& {Penprase},
  B.~E. 1995, \apj, 453, 256

\bibitem[{{Landi} {et~al.}(2013){Landi}, {Young}, {Dere}, {Del Zanna}, \&
  {Mason}}]{Chianti7.1}
{Landi}, E., {Young}, P.~R., {Dere}, K.~P., {Del Zanna}, G., \& {Mason}, H.~E.
  2013, \apj, 763, 86

\bibitem[{{Le{\'o}n-Tavares} {et~al.}(2012){Le{\'o}n-Tavares}, {Valtaoja},
  {Giommi}, {Polenta}, {Tornikoski}, {L{\"a}hteenm{\"a}ki}, {Gasparrini}, \&
  {Cutini}}]{LeonTavares2012}
{Le{\'o}n-Tavares}, J., {Valtaoja}, E., {Giommi}, P., {et~al.} 2012, \apj, 754,
  23

\bibitem[{{Liang} {et~al.}(1997){Liang}, {Dickey}, {Moorey}, \&
  {Ekers}}]{Liang1997}
{Liang}, H., {Dickey}, J.~M., {Moorey}, G., \& {Ekers}, R.~D. 1997, \aap, 326,
  108

\bibitem[{{Meynet} {et~al.}(2001){Meynet}, {Arnould}, {Paulus}, \&
  {Maeder}}]{Wolf-Rayet-isotopes}
{Meynet}, G., {Arnould}, M., {Paulus}, G., \& {Maeder}, A. 2001, \ssr, 99, 73

\bibitem[{{Million} {et~al.}(2011){Million}, {Werner}, {Simionescu}, \&
  {Allen}}]{Million-Virgo-Metals}
{Million}, E.~T., {Werner}, N., {Simionescu}, A., \& {Allen}, S.~W. 2011,
  \mnras, 418, 2744

\bibitem[{{Moustakas} \& {Kennicutt}(2006)}]{Moustakas2006}
{Moustakas}, J., \& {Kennicutt}, Jr., R.~C. 2006, \apjs, 164, 81

\bibitem[{{Nesterov} {et~al.}(1995){Nesterov}, {Lyuty}, \&
  {Valtaoja}}]{Nesterov1995}
{Nesterov}, N.~S., {Lyuty}, V.~M., \& {Valtaoja}, E. 1995, \aap, 296, 628

\bibitem[{{Peterson} \& {Fabian}(2006)}]{Peterson-CoolCore-Spectra}
{Peterson}, J.~R., \& {Fabian}, A.~C. 2006, \physrep, 427, 1

\bibitem[{{Pettini} \& {Bowen}(2001)}]{Pettini2001}
{Pettini}, M., \& {Bowen}, D.~V. 2001, \apj, 560, 41

\bibitem[{{Reid} {et~al.}(2009){Reid}, {Menten}, {Zheng}, {Brunthaler},
  {Moscadelli}, {Xu}, {Zhang}, {Sato}, {Honma}, {Hirota}, {Hachisuka}, {Choi},
  {Moellenbrock}, \& {Bartkiewicz}}]{Dist-parall}
{Reid}, M.~J., {Menten}, K.~M., {Zheng}, X.~W., {et~al.} 2009, \apj, 700, 137

\bibitem[{{Riffel} {et~al.}(2006){Riffel}, {Rodr{\'{\i}}guez-Ardila}, \&
  {Pastoriza}}]{Riffel2006}
{Riffel}, R., {Rodr{\'{\i}}guez-Ardila}, A., \& {Pastoriza}, M.~G. 2006, \aap,
  457, 61

\bibitem[{{Rogers} {et~al.}(2007){Rogers}, {Dudevoir}, \& {Bania}}]{Rogers2007}
{Rogers}, A.~E.~E., {Dudevoir}, K.~A., \& {Bania}, T.~M. 2007, \aj, 133, 1625

\bibitem[{{Rood} {et~al.}(2007){Rood}, {Bania}, \& {Balser}}]{Rood-He3-140ft}
{Rood}, R.~T., {Bania}, T.~M., \& {Balser}, D.~S. 2007, {An Old Dog's Last
  Hunt: The Last Observations of the NRAO Green Bank 140 Foot Radio Telescope},
  ed. F.~J. {Lockman}, F.~D. {Ghigo}, \& D.~S. {Balser} (the National Radio
  Astronomy Observatory, Charlottesville, Virginia), 451

\bibitem[{{Sarazin}(2013, private communication)}]{Sarazin-priv-2013}
{Sarazin}, C.~L. 2013, private communication

\bibitem[{{Seaton}(2005)}]{OpacityProject}
{Seaton}, M.~J. 2005, \mnras, 362, L1

\bibitem[{{Shabaev} {et~al.}(1995){Shabaev}, {Shabaeva}, \&
  {Tupitsyn}}]{Shabaev1995}
{Shabaev}, V.~M., {Shabaeva}, M.~B., \& {Tupitsyn}, I.~I. 1995, \pra, 52, 3686

\bibitem[{{Stone}(2005)}]{Stone2005}
{Stone}, N.~J. 2005, Atomic Data and Nuclear Data Tables, 90, 75

\bibitem[{{Sunyaev} \& {Docenko}(2007)}]{SunyaevDocenko2007}
{Sunyaev}, R.~A., \& {Docenko}, D.~O. 2007, Astronomy Letters, 33, 67

\bibitem[{{Sunyaev} {et~al.}(2003){Sunyaev}, {Norman}, \&
  {Bryan}}]{Sunyaev-ICMturb-2003}
{Sunyaev}, R.~A., {Norman}, M.~L., \& {Bryan}, G.~L. 2003, Astronomy Letters,
  29, 783

\bibitem[{{Syunyaev} \& {Churazov}(1984)}]{SunyaevChurazov1984}
{Syunyaev}, R.~A., \& {Churazov}, E.~M. 1984, Soviet Astronomy Letters, 10, 201

\bibitem[{{Trippe} {et~al.}(2011){Trippe}, {Krips}, {Pi{\'e}tu}, {Neri},
  {Winters}, {Gueth}, {Bremer}, {Salome}, {Moreno}, {Boissier}, \&
  {Fontani}}]{Trippe2011}
{Trippe}, S., {Krips}, M., {Pi{\'e}tu}, V., {et~al.} 2011, \aap, 533, A97

\bibitem[{{Tully} \& {Fisher}(1977)}]{TullyFisher1977}
{Tully}, R.~B., \& {Fisher}, J.~R. 1977, \aap, 54, 661

\bibitem[{{Weedman} {et~al.}(2005){Weedman}, {Hao}, {Higdon}, {Devost}, {Wu},
  {Charmandaris}, {Brandl}, {Bass}, \& {Houck}}]{Weedman2005}
{Weedman}, D.~W., {Hao}, L., {Higdon}, S.~J.~U., {et~al.} 2005, \apj, 633, 706

\bibitem[{{Werner} {et~al.}(2010){Werner}, {Simionescu}, {Million}, {Allen},
  {Nulsen}, {von der Linden}, {Hansen}, {B{\"o}hringer}, {Churazov}, {Fabian},
  {Forman}, {Jones}, {Sanders}, \& {Taylor}}]{Werner-Virgo-2010}
{Werner}, N., {Simionescu}, A., {Million}, E.~T., {et~al.} 2010, \mnras, 407,
  2063

\bibitem[{{Zhang} \& {Sampson}(2000)}]{ZhangSampson2000}
{Zhang}, H.~L., \& {Sampson}, D.~H. 2000, \pra, 61, 022707 (ZS00)

\bibitem[{{Zhang} \& {Sampson}(2001)}]{ZhangSampson2001}
---. 2001, \mnras, 322, 433 (ZS00)

\end{thebibliography}

\end{document}